\RequirePackage{fix-cm} 
\documentclass[a4paper, twoside, reqno, 12pt, dvips]{amsart}

\usepackage{fixltx2e}     

\usepackage{a4wide}

\usepackage{amsmath,amsfonts,amssymb}
\usepackage{latexsym}

\usepackage[latin1]{inputenc}
\usepackage{graphicx}

\usepackage{psfrag}
\usepackage{subfig}
\usepackage{indentfirst}

\usepackage{xspace} 

\usepackage{color}
\definecolor{oneblue}{rgb}{0,0.0,0.75}
\usepackage[colorlinks,
            urlcolor=oneblue,
            linkcolor=oneblue,
            citecolor=oneblue,
            bookmarksopen=false,
            pagebackref]{hyperref}

\usepackage[usenames,dvipsnames]{pstricks}
\usepackage{pst-plot}           
\usepackage{pstricks-add}       
\usepackage{epsfig}

\newtheorem{proposition}{Proposition}
\newtheorem{theorem}{Theorem}

\newtheorem{remark}{Remark}
\newtheorem{definition}{Definition}

\newcommand{\od}[2]{\frac{d#1}{d#2}}
\newcommand{\pd}[2]{\frac{\partial#1}{\partial#2}}
\newcommand{\set}[1]{\left\{ #1 \right\}}
\newcommand{\abs}[1]{\left|#1\right|}
\newcommand{\area}{\mathop{\mathrm{area}}}
\newcommand{\sign}{\mathop{\mathrm{sign}}}
\newcommand{\norm}[1]{\left|\left|#1\right|\right|}
\newcommand{\vol}{\mathop{\mathrm{vol}}}
\newcommand{\dx}{\partial_x}
\newcommand{\TV}{\mathop{\mathrm{TV}}}

\def\m{\mu^2}
\def\u{\vec{u}}
\def\n{\vec{n}}
\def\x{\vec{x}}
\def\w{w}

\def\Om{\Omega}
\def\grad{\nabla}
\def\div{\nabla\cdot}
\def\eps{\varepsilon}
\def\A{\mathbb{A}}
\def\R{\mathbb{R}}

\def\Rc{\mathcal{R}}
\def\F{\mathcal{F}}
\def\T{\mathcal{T}}
\def\L{\mathcal{L}}
\def\O{\mathcal{O}}
\def\N{\mathcal{N}}
\def\S{\mathcal{S}}
\def\rit{\mathbb{R}}
\def\zit{\mathbb{Z}}
\def\nit{\mathbb{N}}
\def\un{\underline}
\def\uvl{\underline{w}}
\def\VOLNA{\textsl{VOLNA }}
\def\cfd{Computational Fluid Dynamics}

\newcommand{\cqfd}
{%
\mbox{}%
\nolinebreak%
\hfill%
\rule{2mm}{2mm}%
\medbreak%
\par%
}

\numberwithin{equation}{section}

\begin{document}

\title[The VOLNA code for the numerical modelling of tsunami waves]{The VOLNA code for the numerical modelling of tsunami waves: generation, propagation and inundation}

\author[D. Dutykh]{Denys Dutykh}
\address{Centre de Mathmatiques et de Leurs Applications, ENS Cachan and CNRS, UniverSud, 61 avenue du
  President Wilson, F-94235 Cachan Cedex, and LRC MESO, ENS Cachan, CEA DAM DIF\\
  Now at Universit de Savoie, Laboratoire de Mathmatiques LAMA - UMR 5127,
Campus Scientifique, 73376 Le Bourget-du-Lac Cedex, France}
\email{Denys.Dutykh@univ-savoie.fr}
\urladdr{http://www.lama.univ-savoie.fr/~dutykh/}

\author[R. Poncet]{Rapha\"{e}l Poncet}
\address{Centre de Mathmatiques et de Leurs Applications,\\ ENS Cachan and CNRS, UniverSud, 61 avenue du
  President Wilson, F-94235 Cachan Cedex, and LRC MESO, ENS Cachan, CEA DAM DIF\\ Now at CEA, DAM, DIF, F-91297 Arpajon, France}
\email{raphael.poncet@gmail.com}

\author[F. Dias]{Fr\'{e}d\'{e}ric Dias$^*$}
\thanks{$^*$ Corresponding author}
\address{Centre de Mathmatiques et de Leurs Applications, ENS Cachan and CNRS, UniverSud, 61 avenue du President Wilson, F-94235 Cachan Cedex, and LRC MESO, ENS Cachan, CEA DAM DIF \\ On leave at University College Dublin, School of Mathematical Sciences, Belfield, Dublin 4, Ireland}
\email{Frederic.Dias@cmla.ens-cachan.fr}
\urladdr{http://www.cmla.ens-cachan.fr/Membres/dias}

\begin{abstract}
A novel tool for tsunami wave modelling is presented. This tool has the potential of being used for operational purposes: indeed, the numerical code \VOLNA is able to handle the complete life-cycle of a tsunami (generation, propagation and run-up along the coast). The algorithm works on unstructured triangular meshes and thus can be run in arbitrary complex domains. This paper contains the detailed description of the finite volume scheme implemented in the code. The numerical treatment of the wet/dry transition is explained. This point is crucial for accurate run-up/run-down computations. Most existing tsunami codes use semi-empirical techniques at this stage, which are not always sufficient for tsunami hazard mitigation. Indeed the decision to evacuate inhabitants is based on inundation maps which are produced with this type of numerical tools. We present several realistic test cases that partially validate our algorithm. Comparisons with analytical solutions and experimental data are performed. Finally the main conclusions are outlined and the perspectives for future research presented.
\end{abstract}


\maketitle

\tableofcontents


\section{Introduction}\label{sec:intro}

After the 2004 Boxing Day tsunami \cite{Syno2006} and the 2011 Honshu, Japan tsunami, there is no need to explain the importance of research on tsunami waves. One of the primary objectives in this field consists in establishing and developing Tsunami Warning Systems (TWS) \cite{Tatehata1997, Titov2005} and inundation maps. This task is non trivial as explained by Synolakis \cite{Synolakis2005}:
\begin{quote}
For reference, the United States and Japan took more than 20 years to develop \textit{validated numerical models} to predict tsunami evolution. And it took the US National Oceanic and Atmospheric Administration 30 years to fully develop its bottom-pressure recorders, which have been reliably detecting tsunamis for the past ten years.
\end{quote}
After the Boxing Day tsunami, while developing their own national and regional capabilities, countries in the Indian Ocean and the Caribbean Sea have asked the PTWC (Pacific Tsunami Warning Center) to act as their interim warning center. India and Australia now have
fully working national centers, while the National Oceanic and Atmospheric Administration of the U.S. has assisted
both with instrumentation and the sophisticated forecast technology used in the Pacific. Europe however is trying to reinvent the early warning wheel. As a result, the Mediterranean remains the only world sea unprotected by any warning system. The 2011 Honshu tsunami also
showed that the tsunami community did not do enough to anticipate future events, even in Japan which is arguably the most tsunami-ready nation in the world.

The mathematical modelling and computation of propagating tsunami waves play an important r\^ole in TWS. Precision and robustness of the algorithm will affect performance and reliability of the whole system.

The importance of tsunami generation modelling is often underestimated by the scientific community. During several years the research of our group was focused on this topic and interesting results were obtained \cite{Dutykh2006, Dutykh2007a, ddk, Kervella2007, Dutykh2007b, Dutykh2008, Dutykh2010a, Dutykh2010d}. We tried to incorporate some recent developments \cite{Dutykh2006} from this field into the \VOLNA code.

The recent events in Japan should convince the scientific community of the urgency to complete inundation maps. These are maps that show the extent of possible tsunami flooding from hypothetical earthquakes in their vicinity. Even in the U.S., only California has completed its mapping efforts. Alaska and Hawaii, the most vulnerable U.S. states, do not have modern tsunami flood maps for all their coastal communities.\footnote{When will we learn, Newsweek, March 13, 2011, by Costas Synolakis}

It is difficult to find a topic in numerical analysis of hyperbolic PDEs which has been studied more than the numerical solution to the Nonlinear Shallow Water Equations (NSWE). The numerical scheme presented in this paper is not completely novel. The discretization methods used in \VOLNA can be found in the modern literature on finite volumes methods \cite{Kroner1997, Barth2004}. The main purpose here is to present a tool for tsunami wave modelling which covers the whole spectrum from generation to inundation. The emphasis is on the technical work which is typical of a numerical analyst and software developer. Tsunami practicioners can then concentrate on the physical aspects of tsunami propagation.

Nowadays, one is facing a somewhat strange situation. On one hand, there are only a few truly operational codes for tsunami wave modelling: MOST, NAMI, ComCot \cite{Imamura1996, Titov1997, Goto1997, Liu1998}. The numerical schemes used in these codes essentially correspond to the state of the art of the eighties. On the other hand, there is a plethora of NSWE codes developed in academic environments \cite{Glaister1988, Casulli1990, Toro1992, Bermudez1998, Anastasiou1999, Vazquez-Cendon1999, Alcrudo2005, Benkhaldoun2006, Garcia-Navarro2000, Gallouet2003, Audusse2004, Kim2007, George2006, George2006a, Zhou2002, Causon2000, Noelle2006, Castro2005, Wei2006, George2008, DeKaKa, CBB1, CBB2} -- see \cite{Barthelemy2004} for a review of nonlinear shallow water theories for coastal waves. These codes use modern numerical methods but most of them have not been developed to satisfy the needs of tsunami operational research. This is why we had the idea to develop \VOLNA. We tried to combine modern numerical techniques for hyperbolic systems with real world application-oriented design. The \VOLNA code can be run efficiently in realistic environments. It was shown that natural coasts tend to have fractal forms \cite{Sapoval2004}. Hence, unstructured meshes are a natural choice in this type of situations. The first demonstration of the applicability of VOLNA to real situations was provided by Poncet et al. \cite{PoncetCanada2010} in their study of tsunamis generated by landslides in the St. Lawrence estuary.

The paper is organised as follows. In Section \ref{sec:model} the physical context of the study and the motivation for the choice of the mathematical model are presented. Section \ref{sec:discret} contains a detailed description of the numerical method implemented in the \VOLNA code. In Section \ref{sec:results} we show computations which validate and illustrate the capabilities of \VOLNA. Finally the main conclusions are outlined and the perspectives for future research presented in Section \ref{sec:concl}.


\section{Physical context and mathematical model}\label{sec:model}

\begin{figure}
  \centering
  \includegraphics[width=0.8\textwidth]{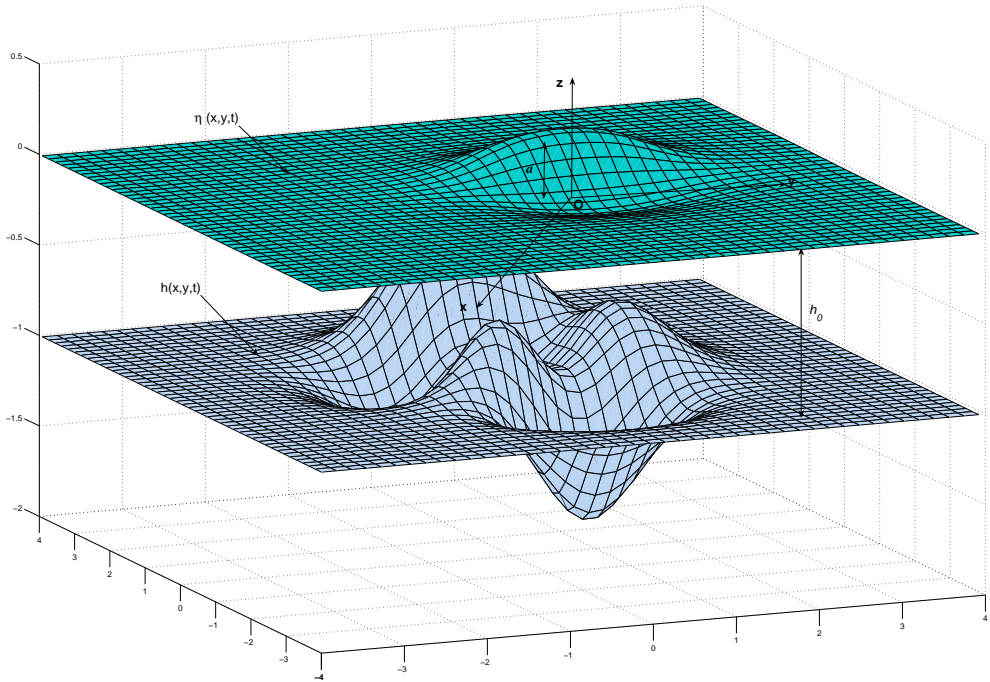}
  \caption{Sketch of the fluid domain.}
  \label{fig:sketch}
\end{figure}

In this study we focus on long wave propagation over realistic bathymetry. A sketch of the physical problem under investigation is given on \figurename~\ref{fig:sketch}. Let us explain the main assumptions and the domain of applicability of the \VOLNA code.

First we introduce some characteristic lengths. We denote by $a_0$ the typical wave amplitude, by $h_0$ the average depth and by $\ell$ the characteristic wave length. Several dimensionless numbers can be built from these three quantities, but traditionally one introduces
the following two:
\begin{equation*}
  \eps = \frac{a_0}{h_0}, \quad \m = \left(\frac{h_0}{\ell}\right)^2.
\end{equation*}
The first parameter $\eps$ measures the wave nonlinearity ($\eps \ll 1$ means than nonlinearity is weak) while the second parameter $\m$ quantifies the importance of dispersive effects ($\mu \ll 1$ means than dispersion is weak). Taking a typical megatsunami offshore with 
roughly
\begin{equation*}
  a_0 \simeq 0.5\;m, \quad h_0 \simeq 4\;km, \quad \ell \simeq 100\;km
\end{equation*}
yields $\eps = 1.25\times 10^{-4}$ for the nonlinearity parameter and $\m = 1.6\times 10^{-3}$ for the dispersion parameter. Both are weak. Using asymptotic expansions in the small parameters $\eps\ll 1$ and $\m\ll 1$, one can derive Serre-type equations \cite{Serre1953, Peregrine1967, Madsen03, Lynett, Dutykh2007, Lannes2009, Dias2010, ChazelLannes2010, Bonneton2011, Clamond2009, Carter2011, Dutykh2011a}. The effect of dispersion on tsunamis has been investigated recently \cite{Tkalich2007, Ioualalen2007, MFS2008}. Due to frequency dispersion, longer and higher waves travel faster and separate from the shorter and smaller waves, leading to a decrease of tsunami height. It is close to the shore that dispersion might play a r\^ole. Here the short waves may have a local additional effect on wave impact on coastal structures, but they hardly play any role for the runup and inundation caused by the main and much longer tsunami. Therefore the consequences of neglecting dispersive effects are probably not very important from a practical point of view. Moreover no operational codes based on the Serre equations exist as of today. The GEOWAVE code, which combines TOPICS and FUNWAVE, is still too expensive from a computational point of view to be used as a truly operational code. Neglecting the dispersive effects yields the classical Nonlinear Shallow Water Equations (NSWE):
\begin{align}
  H_t + \div (H\u) &= 0, \label{eq:gov1} \\
  (H\u)_t + \div\left(H\u\otimes\u + \frac{g}{2}H^2\right) &= gH\grad h, \label{eq:gov2}
\end{align}
where $H = h + \eta$ is the total water depth and $\u = (u,v) (\x,t)$ is the depth-averaged horizontal velocity. Traditionally, $g$ denotes the acceleration due to the gravity and $h(\x,t)$ describes the bathymetry.
\begin{remark}
The bathymetry $h(\x,t)$ is allowed to be time-dependent. It is important for the problem of tsunami generation by underwater earthquakes, submarine landslides, etc. The coupling with seismology is done through this function. Namely, various simplified earthquake models \cite{Dutykh2006, Kervella2007, Dutykh2007b, Dutykh2008, Dutykh2010a, Dutykh2010d} provide the seabed displacements which are then transmitted to the ocean layer.
\end{remark}

In this study, the NSWE (\ref{eq:gov1}) and (\ref{eq:gov2}) are chosen to model tsunami generation, propagation and run-up/run-down. It is computationally advantageous to have a uniform model for all stages of tsunami life since many technical problems are thus avoided. The validity of the NSWE for tsunami generation was already examined in our previous study \cite{Kervella2007}, where an excellent performance of this model was shown for nondispersive long waves. In the present paper we show in Section \ref{sec:results} the ability of the NSWE to model the run-up/run-down process. For this purpose, comparisons with a laboratory experiment are performed. Thus, the chosen \textit{complete} approach to tsunami wave modelling is very attractive from both the operational and research viewpoints.

The governing equations (\ref{eq:gov1}) and (\ref{eq:gov2}) have nice mathematical properties. In particular, this system is strictly hyperbolic provided that $H > 0$. This property will be used extensively in the construction of the numerical scheme (see Section \ref{sec:discret}).

Let us discuss the eigensystem of the advective flux. First, we introduce conservative variables and rewrite the governing equations as a system of conservation laws:
\begin{equation}\label{eq:diffsys}
  \pd{\w}{t} + \div\F(\w) = \S (\w),
\end{equation}
where the following notation was introduced:
\begin{equation*}
  \w(x,t):\R^2\times\R^+\mapsto \R^3, \qquad \w = (w_1, w_2, w_3) = (H, Hu, Hv),
\end{equation*}
\begin{equation*}
  \F (\w) = \begin{pmatrix}
    Hu & Hv \\
    Hu^2 + \frac{g}{2}H^2 & Huv \\
    Huv & Hv^2 + \frac{g}{2}H^2 \\
  \end{pmatrix} =
  \begin{pmatrix}
    w_2 & w_3 \\
    \frac{w_2^2}{w_1} + \frac{g}{2}w_1^2 & \frac{w_2w_3}{w_1} \\
    \frac{w_2w_3}{w_1} & \frac{w_3^2}{w_1} + \frac{g}{2}w_1^2 \\
  \end{pmatrix},
  \S(\w) = \begin{pmatrix}
    0 \\
    gH\pd{h}{x} \\
    gH\pd{h}{y} \\
  \end{pmatrix}.
\end{equation*}
After projecting the flux $\F(\w)$ in the normal direction $\n = (n_x, n_y)$ (face normal), one can compute the Jacobian matrix $\A_n$. Its expression in physical variables has the following form:
\begin{equation*}
  \A_n = \pd{\bigl(\F(\w)\cdot\n\bigr)}{\w} =
  \begin{pmatrix}
    0 & n_x & n_y \\
    -u u_n + gHn_x & u_n + u n_x & u n_y \\
    -v u_n + gHn_y & v n_x & u_n + v n_y \\
  \end{pmatrix},
\end{equation*}
where $u_n = u n_x + v n_y$ is the velocity vector projected on $\n$. The Jacobian matrix $\A_n$ has three distinct eigenvalues:
\begin{equation}\label{eq:eigen}
  \lambda_1 = u_n - c, \quad \lambda_2 = u_n, \quad \lambda_3 = u_n + c,
\end{equation}
where $c = \sqrt{gH}$ is the speed of gravity waves in the limit of infinite wavelength. This quantity plays the same r\^ole as the sound speed in compressible fluid mechanics. It is now obvious that the system (\ref{eq:gov1}), (\ref{eq:gov2}) is strictly hyperbolic provided that $H > 0$. The eigenstructure of the Jacobian matrix $\A_n$ is fundamental for constructing the numerical flux function (see Section \ref{sec:firstorder}) and thus, upwinding the discrete solution.


\section{Discretization procedure}\label{sec:discret}

In this study we selected the most natural numerical method for this type of equations. Finite volume (FV) methods are a class of discretization schemes that have proven highly successful in solving numerically a wide class of systems of conservation laws. These systems often come from compressible fluid dynamics. In electromagnetism, for example, discontinuous Galerkin methods have proven to be more efficient \cite{Cockburn2004}. When compared to other discretization methods such as finite elements or finite differences, the primary advantages of FV methods are robustness, applicability on general unstructured meshes, and the intrinsic local conservation properties. Hence, with this type of discretization, mass, momentum and total energy are conserved exactly, at least in the absence of source terms and appropriate boundary conditions.

In order to solve numerically the system of balance laws (\ref{eq:gov1}), (\ref{eq:gov2}) one uses again the conservative form of governing equations (\ref{eq:diffsys}). System (\ref{eq:diffsys}) should be provided with an initial condition
\begin{equation}\label{eq:initialcond}
  \w(\x,0) = \w_0(\x), \quad \x = (x,y) \in \Omega
\end{equation}
and appropriate boundary conditions. The implementation of different boundary conditions will be discussed below (see Section \ref{sec:bound}).

\subsection{First order scheme}\label{sec:firstorder}

\begin{figure}[htbp]
  \centering
  \psfrag{O}{$O$}
  \psfrag{K}{$K$}
  \psfrag{M}{$\partial K$}
  \psfrag{n}{$\n_{KL}$}
  \psfrag{L}{$L$}
  \includegraphics[width=5cm]{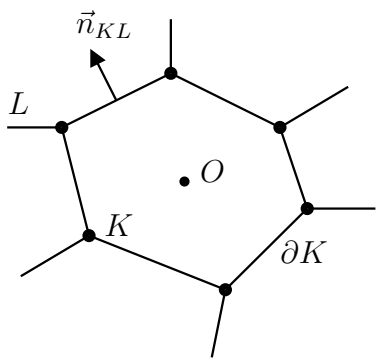}
  \caption[An example of control volume $K$.]{An example of control volume $K$ with barycenter $O$. The normal pointing from $K$ to $L$ is denoted by $\n_{KL}$.}
  \label{fig:controlvol}
\end{figure}

The computational domain $\Omega\subset\R^2$ is triangulated into a set of non overlapping control volumes that completely cover the domain. Let $\T$ denote a tesselation of the domain $\Omega$ with control volume $K$ such that
\begin{equation*}
  \cup_{K\in\T} \bar{K} = \bar{\Omega}, \quad \bar{K} := K \cup \partial K.
\end{equation*}
For two distinct control volumes $K$ and $L$ in $\T$, the intersection is an edge with oriented normal $\n_{KL}$ or else a vertex. We need to introduce the following notation for the neighbourhood of $K$:
\begin{equation*}
  \N(K) := \set{L\in\T: \area(K\cap L) \neq 0},
\end{equation*}
a set of all control volumes $L$ which share an edge in 2D or a face in 3D with the given volume $K$. In this study, we denote by $\vol(\cdot)$ and $\area(\cdot)$ the area and length respectively.

\begin{figure}[htbp]
  \centering
  \psfrag{a}{Storage location}
  \psfrag{b}{Control volume}
  \includegraphics[width=7cm]{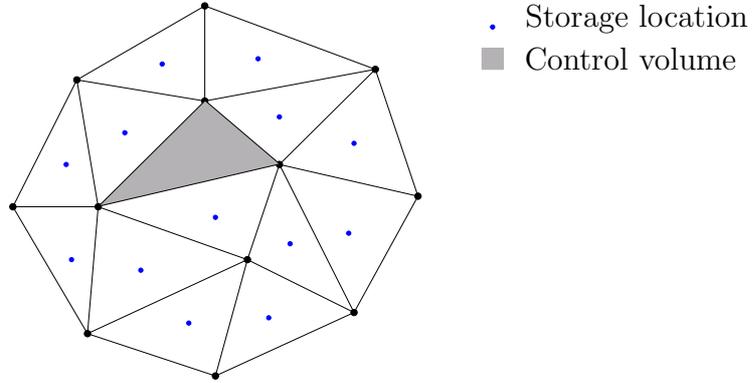}
  \caption[Illustration for cell-centered finite volume method]{Illustration for cell-centered finite volume method}
  \label{fig:cell}
\end{figure}

The choice of control volume tesselation is flexible in the FV method. In the present study we selected the cell-centered approach (see \figurename~\ref{fig:cell}), which means that degrees of freedom are associated to cell barycenters.

The first steps in FV methods are classical. One starts by integrating equation (\ref{eq:diffsys}) on the control volume $K$ shown in \figurename~\ref{fig:controlvol} and one applies Gauss-Ostrogradsky theorem for advective and diffusive fluxes. Then, in each control volume, an integral conservation law is imposed:
\begin{equation}\label{eq:conservlaw}
  \od{}{t}\int_{K} \w \;d\Omega + \int_{\partial K}\F(\w)\cdot\n_{KL} \;d\sigma = \int_{K}\S(\w) \;d\Omega\;
\end{equation}
Physically an integral conservation law states that the rate of change of the total amount of a quantity (for example: mass, momentum, total energy) with density $\w$ in a fixed control volume $K$ is balanced by the flux $\F$ of the quantity through the boundary $\partial K$ and the production of this quantity $\S$ inside the control volume.

The next step consists in introducing the control volume cell average for each $K\in\T$
\begin{equation*}
  \w_K(t) := \frac{1}{\vol(K)}\int_{K} \w(\x,t) \;d\Omega \;.
\end{equation*}
After the averaging step, the FV method can be interpreted as producing a system of evolution equations for cell averages, since
\begin{equation*}
  \od{}{t}\int_{K} \w(\x,t) \;d\Omega = \vol(K)\od{\w_K}{t} \;.
\end{equation*}
Godunov was first \cite{Godunov1959} to pursue and apply these ideas to the discretization of the gas dynamics equations.

However, the averaging process implies piecewise constant solution representation in each control volume with value equal to the cell average. The use of such a representation makes the numerical solution multivalued at control volume interfaces. Thereby the calculation of the fluxes $\int_{\partial K}(\F(\w)\cdot\n_{KL}) \;d\sigma$ at these interfaces is ambiguous. A fundamental aspect of FV methods is the idea of substituting the true flux at interfaces by a numerical flux function
\begin{equation*}
  \left.\bigl(\F(\w)\cdot\n\bigr)\right|_{\partial K\cap\partial L} \longleftarrow \Phi(\w_K, \w_L; \n_{KL}) : \R^3\times\R^3 \mapsto \R^3 \;,
\end{equation*}
a Lipschitz continuous function of the two interface states $\w_K$ and $\w_L$. The key ingredient is the choice of the numerical flux function $\Phi$. In general this function is calculated as an exact or even better approximate local solution of the Riemann problem posed at these interfaces. In the present study we implemented several numerical fluxes (HLL, HLLC, FVCF) described below.

Any numerical flux is assumed to satisfy the following properties:
\begin{description}
\item[Conservation.] This property ensures that fluxes from adjacent control volumes sharing an interface exactly cancel when summed. This is achieved if the numerical flux function satisfies the identity
  \begin{equation*}
    \Phi(\w_K,\w_L; \n_{KL}) = - \Phi(\w_L,\w_K; \n_{LK}).
  \end{equation*}
\item[Consistency.] Consistency is obtained when the numerical flux with identical state arguments reduces to the true flux of the same state, i.e.
  \begin{equation*}
    \Phi(\w,\w; \n) = (\F(\w)\cdot\n)(\w).
  \end{equation*}
\end{description}

In the following paragraphs \ref{sec:vffc} -- \ref{sec:hllc} we give several examples of numerical flux functions $\Phi$ 
which were implemented in the \VOLNA code. These choices are justified by efficiency, clarity and personal preferences of the authors. However, we do not impose them and a final user can easily implement his favourite numerical flux function.

\subsubsection{FVCF approach}\label{sec:vffc}

First we describe the scheme called Finite Volumes with Characteristic Flux (FVCF) and proposed by Ghidaglia et al. in \cite{Ghidaglia1995, Ghidaglia1996, Ghidaglia2001}.

Consider a general system of conservation laws in 1D that can be written as follows:
\begin{equation}\label{1.24}
        \frac{\partial w}{\partial t}+\frac{\partial f(w)}{\partial x}=0 \,,
\end{equation}
where $w \in \rit^m$ and $f~:\rit^m\mapsto \rit^m$. We denote by $A(w)$ 
the Jacobian matrix $\frac{\partial f(w)}{\partial w}$ and we deal 
with the case where (\ref{1.24}) is \textit{smoothly hyperbolic}, that is 
to say: for every $w$ there exists a smooth basis $(r_1(w),\ldots ,r_m(w))$ 
of $\rit^m$ consisting of eigenvectors of $A(w)$. That is $\exists \lambda_k(w) 
\in \rit \mbox{ such that } A(w)r_k(w)=\lambda_k(w) r_k(w)$. It is then 
possible to construct $(l_1(w),\ldots ,l_m(w))$ such that 
$^tA(w)l_k(w) = \lambda_k(w) l_k(w)$ and $l_k(w)\cdot r_p(w) = \delta_{k,p}$.

Let ${\rit} =\cup_{j\in \zit} [x_{j-1/2}, x_{j+1/2}]$ be a 1D mesh. The goal is to discretize (\ref{1.24}) by a FV method. We set $\Delta x_j \equiv x_{j+1/2} - x_{j-1/2}$, $\Delta t_n \equiv {t_{n+1}}-{t_n}$ (we also have ${\rit}_+ = \cup_{n\in \nit} [t_n, t_{n+1}]$) and
$$ 
\tilde w^n_j \equiv \frac{1}{\Delta x_j} \int^{x_{j+1/2}}_{x_{j-1/2}} w(x,t_n) \,dx\,, \quad
\tilde f^n_{j+1/2}\equiv \frac{1}{\Delta t_n} \int_{t_n}^{t_{n+1}}f(w(x_{j+1/2},t)) \,dt\,.
$$ 
With these notations, we deduce from (\ref{1.24}) the {\it exact} relation:
\begin{equation}\label{1.26}
        \tilde w^{n+1}_j=\tilde w^n_j-\frac{\Delta t_n}{\Delta x_j} \left(
        \tilde f^{n}_{j+1/2} - \tilde f^{n}_{j-1/2}\right)\,.
\end{equation}
Since the $(\tilde f^n_{j+1/2})_{j\in\zit}$ cannot be expressed in terms of the $(\tilde w^n_{j})_{j\in\zit}$, one has to make an approximation. In order to keep a compact stencil, it is more efficient to use a three point scheme: the physical flux $\tilde f^n_{j+1/2}$ is approximated by a numerical flux $g_j^n(w^n_j,w^n_{j+1})$. Let us show how this flux is constructed here. Since $A(w) \frac{\partial w}{\partial t} = \frac{\partial f(w)}{\partial t}$ we observe that according to (\ref{1.24})
\begin{equation}\label{1.25}
        \frac{\partial f(w)}{\partial t} + A(w) \frac{\partial f(w)}{\partial x}=0\,.
\end{equation}

This shows that the flux $f(w)$ is advected by $A(w)$ like $w$. The numerical flux $g_j^n(w^n_j,w^n_{j+1})$ represents the flux at an interface. Using a mean value $\mu_{j+1/2}^n$ of $w$ at this interface, we replace (\ref{1.25}) by the linearization:
\begin{equation}\label{1.27}
        \frac{\partial f(w)}{\partial t} + A(\mu_{j+1/2}^n) \frac{\partial f(w)}{\partial x}=0\,.
\end{equation}

It follows that, defining the $k$-th characteristic flux component to be $f_k(w)\equiv l_k(\mu_{j+1/2}^n)\cdot f(w)$, one has
\begin{equation}\label{1.28}
        \frac{\partial f_k(w)}{\partial t}+\lambda_k(\mu_{j+1/2}^n) \frac{\partial
        f_k(w)}{\partial x}=0\,.
\end{equation}
This linear equation can be solved explicitly:
\begin{equation}\label{1.29}
        f_k(w)(x,t)=f_k(w)(x-\lambda_k(\mu_{j+1/2}^n)(t-t_n),t_n)\,.
\end{equation}
From this equation it is then natural to introduce the following definition.

\begin{definition}\label{def1} 
        For the conservative system (\ref{1.24}), at the interface between the two cells $[x_{j-1/2}, x_{j+1/2}]$ and $[x_{j+1/2}, x_{j+3/2}]$, the characteristic flux $g^{CF}$ is defined by the following formula for $k\in\{1,\ldots,m\}$ : \\
        $\left(\mbox{we take }\mu_{j+1/2}^n\equiv \bigl(\Delta x_j w_j^n +\Delta x_{j+1}
        w^n_{j+1}\bigr)/\bigl(\Delta x_j +\Delta x_{j+1}\bigr)\right)$
        \begin{equation}
                \nonumber
                l_k(\mu_{j+1/2}^n)\cdot g^{CF,n}_j(w^n_j,w^n_{j+1}) =
                l_k(\mu_{j+1/2}^n)\cdot f(w^n_j)\,,\mbox{ when } \lambda_k(\mu_{j+1/2}^n)>0\,,
        \end{equation}
        \begin{equation}\label{1.30}
                l_k(\mu_{j+1/2}^n)\cdot g^{CF,n}_j(w^n_j,w^n_{j+1}) =
                l_k(\mu_{j+1/2}^n)\cdot f(w^n_{j+1})\,,\mbox{ when }
                \lambda_k(\mu_{j+1/2}^n)<0\,,\,
        \end{equation}
        \begin{equation*}
                l_k(\mu_{j+1/2}^n)\cdot g^{CF,n}_j(w^n_j,w^n_{j+1}) =
                l_k(\mu_{j+1/2}^n)\cdot \left(\frac{f(w^n_{j+1})+f(w^n_{j})}{2}\right)\,,
        \end{equation*}
        when $\lambda_k(\mu_{j+1/2}^n)=0$.
\end{definition}

\begin{remark}
At first glance, the derivation of (\ref{1.25}) from (\ref{1.24}) is only valid for continuous solutions since $A(w) \frac{\partial f(w)}{\partial x}$ is a non conservative product. In fact, equation (\ref{1.25}) can be justified even in the case of shocks as proved in \cite{Ghidaglia1998}. Let us briefly recall here the key point. Assuming that the solution undergoes a discontinuity along a family of disjoint curves, we can focus on one of these curves that we parameterize by the time variable $t$. Hence, locally, on each side of this curve, $w(x,t)$ is smooth and jumps across the curve $x=\Sigma(t)$. The Rankine-Hugoniot condition implies that $f(w(x,t))-\sigma(t)w(x,t)$, where $\sigma(t)\equiv \frac{d\Sigma(t)}{dt}$, is smooth across the discontinuity curve and therefore $A(w) \frac{\partial f(w)}{\partial x}$ can be defined as $A(w) \frac{\partial f(w)}{\partial x}\equiv A(w)       \frac{\partial (f(w)-\sigma w)}{\partial x}+\sigma\frac{\partial f(w)}{\partial x}\,.$
\end{remark}

\begin{proposition}
Formula (\ref{1.30}) can be written as follows: $g^{CF,n}_j(w^n_j,w^n_{j+1}) =  g^{CF}(\mu^n_j; w^n_j, w^n_{j+1})$ where
        \begin{multline}\label{1.31}
                g^{CF}(\mu;v,w) \equiv \sum_{\lambda_k(\mu) < 0} (l_k(\mu)\cdot 
                f(w))r_k(\mu) + \sum_{\lambda_k(\mu) = 0} \left(l_k(\mu)\cdot
        \frac{f(v)+f(w)}{2}\right)r_k(\mu) +\\+ \sum_{\lambda_k(\mu) > 0} 
        (l_k(\mu)\cdot f(v))r_k(\mu) \,.
        \end{multline}
\end{proposition}

{\it Proof.} This comes from the useful identity valid for all vectors $\Phi$ and $\mu$ in $\rit^m$:\\$\Phi=\displaystyle\sum_{k=1}^{k=m} (l_k(\mu)\cdot\Phi)r_k(\mu)$. We also observe that (\ref{1.31}) can be written under the following condensed form:
\begin{equation}\label{1.32}
        g^{CF}(\mu;v,w)=\frac{f(v) + f(w)}{2} - U(\mu;v,w) \frac{f(w) - f(v)}{2}\,,
\end{equation}
where $U(\mu;v,w)$ is the sign of the matrix $A(\mu)$ which is defined by 
$$
        \sign(A(\mu))\Phi=\displaystyle\sum_{k=1}^{m} \sign(\lambda_k)
        (l_k(\mu)\cdot\Phi)r_k(\mu).
$$ 
The form (\ref{1.32}) refers to a numerical flux leading to a flux scheme \cite{Ghidaglia1998}.
\cqfd

\begin{remark}\label{rem2}
        Let us discuss the relation, {\it in the conservative case}, between the characteristic numerical flux $g^{CF}$ and the numerical flux leading to Roe's scheme \cite{Roe1981}. The latter scheme relies on an algebraic property of the continuous flux $f(w)$ which is as follows. It is assumed that for all admissible states $v$ and $w$, there exists a $m \times m$ matrix $A^{ROE}(v,w)$ such that $f(v)-f(w)=A^{ROE}(v,w)(v-w)$ (Roe's identity). Then the numerical flux leading to Roe's scheme is given by:
\begin{equation}\label{1.320}
        g^{ROE}(v,w)=\frac{f(v) + f(w)}{2} - |A^{ROE}(v,w)| \frac{w - v}{2}\,.
\end{equation}
But using Roe's identity, we obtain that
\begin{equation}\label{1.321}
        g^{ROE}(v,w)=\frac{f(v) + f(w)}{2} - \sign(A^{ROE}(v,w)) \frac{f(w) - f(v)}{2}\,,
\end{equation}
which is of the form (\ref{1.32}): Roe's scheme is also a flux scheme. The characteristic flux proposed in this paper is more versatile than Roe's scheme in the sense that it does not rely on an algebraic property of the flux. Hence for complex systems (like those encountered in the context of two phase flows) this scheme is an efficient generalization of Roe's scheme. Moreover, as we shall see below, this scheme has a natural generalization to arbitrary non conservative systems. Finally, the fact that the numerical flux is a linear combination of the two fluxes induces a quite weak dependence on the state $\mu$ which appears in formula (\ref{1.31}), see \cite{Cortes2000}.
\end{remark}

\subsubsection{HLL numerical flux}\label{sec:hll}

\begin{figure}[htbp]
  \centering
  \psfrag{O}{$0$}
  \psfrag{t}{$t$}
  \psfrag{x}{$x$}
  \psfrag{UL}{$w_L$}
  \psfrag{UR}{$w_R$}
  \psfrag{US}{$w^*$}
  \psfrag{SL}{$s_L$}
  \psfrag{SR}{$s_R$}
  \includegraphics[width=6cm]{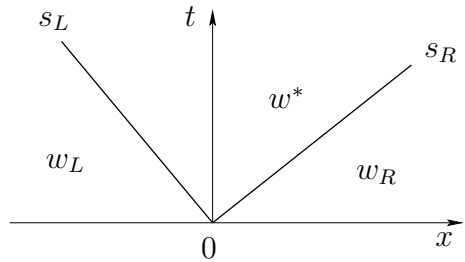}
  \caption{Approximate Riemann fan corresponding to the HLL scheme.}
  \label{fig:HLLfan}
\end{figure}

Now we present another approximate Riemann solver which was proposed by Harten, Lax and van Leer \cite{Harten1983a}. Nowadays this method is known as the HLL scheme. While the exact solution to the Riemann problem contains a large amount of detail, the HLL solver assumes fewer intermediate waves. The simplified Riemann fan is illustrated on Figure \ref{fig:HLLfan}. It consists of two waves separating three constant states.


\begin{figure}
  \centering
  \psfrag{wL}{$w_L$}
  \psfrag{wR}{$w_R$}
  \psfrag{wS}{$w^*$}
  \psfrag{RH}{R.-H.}
  \includegraphics[width=6cm]{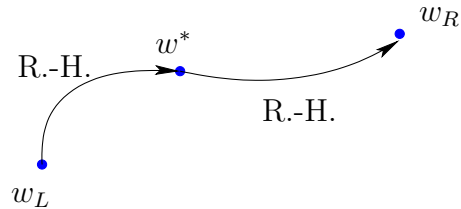}
  \caption{Two states $w_L$ and $w_R$ connected by Rankine-Hugoniot curves represented in the phase space.}
  \label{fig:phaseSpace}
\end{figure}


Consider the following Riemann problem:
\begin{equation}\label{eq:riemann}
  \Rc(w_L, w_R): \left\{
    \begin{array}{l}
      \pd{w}{t} + \pd{F(w)}{x} = 0, \\
      w(x,0) = \left\{
        \begin{array}{l}
          w_L, \quad x < 0, \\
          w_R, \quad x > 0.
        \end{array}
      \right.
    \end{array}
  \right.
\end{equation}

The intermediate state in the approximate Riemann fan will be denoted by $w^*$ and two shock wave speeds are denoted by $s_L$ and $s_R$ respectively (see Figures \ref{fig:HLLfan} and \ref{fig:phaseSpace} for illustration). In order to determine the unknown intermediate state, we write the Rankine-Hugoniot conditions twice:
\begin{equation*}
  \left\{
  \begin{array}{rl}
          s_L(w^* - w_L) &= F^* - F_L, \\
          s_R(w_R - w^*) &= F_R - F^*,
  \end{array}
  \right.
\end{equation*}
where $F_{L,R} := F(w_{L,R})$. It is straightforward to find the solution to this system:
\begin{equation*}
  w^* = \frac{s_Rw_R - s_Lw_L - (F_R - F_L)}{s_R - s_L},
\end{equation*}
\begin{equation}\label{eq:Fstar}
  F^* = F_L + s_L(w^* - w_L) = \frac{s_RF_L - s_LF_R + s_Ls_R(w_R - w_L)}{s_R - s_L}.
\end{equation}
Now we have all the elements to define the numerical flux of the HLL scheme:
\begin{equation*}
  \Phi_{HLL} (w_L, w_R) := \left\{
    \begin{array}{l}
      F_L, \quad s_L \geq 0, \\
      F^*, \quad s_L < 0 \leq s_R, \\
      F_R, \quad s_R < 0.
    \end{array}
  \right.
\end{equation*}

During the presentation of the HLL scheme we missed one important point: how to estimate the wave speeds $s_L$ and $s_R$? The answer is crucial for the overall performance of the scheme. With appropriate choices for the wave speeds $s_L$ and $s_R$, the HLL scheme possesses nice numerical properties. Namely, it satisfies an entropy inequality \cite{Davis1988}, resolves isolated shocks exactly \cite{Harten1983a} and preserves positivity \cite{Einfeldt1991}. In our code we implemented the following choice for $s_L$ and $s_R$ which is motivated by analytical expressions for the Jacobian eigenvalues (\ref{eq:eigen}):
\begin{equation*}
  s_L = \min (u_L - c_L, u^* - c^*), \quad
  s_R = \min (u^* + c^*, u_R + c_R), 
\end{equation*}
where $c_{L,R} := \sqrt{gH_{L,R}}$ is the gravity wave speed for the left and right states and
\begin{equation*}
  u^* = \frac12 (u_L + u_R) + c_L - c_R, \quad
  c^* = \frac12 (c_L + c_R) - \frac14(u_R - u_L).
\end{equation*}
Numerical experiments show that this approximate Riemann solver is very robust with the above choice for the wave speeds \cite{Causon2000, Zhou2002}. One can show that the HLL scheme belongs to the class of flux schemes. Recall that a FV scheme is called a flux scheme if its numerical flux can be written in the following form:
\begin{equation*}
  \Phi = \frac{F(w_L) + F(w_R)}{2} - U(w_L, w_R)\frac{F(w_R) - F(w_L)}{2},
\end{equation*}
where $U(w_L, w_R)$ is some matrix. The robustness of the HLL scheme can be explained by this nice property.

However, the HLL scheme has one important shortcoming: it cannot resolve isolated contact discontinuities. In the next section \ref{sec:hllc} we present another scheme which was designed to remedy this problem.


\subsubsection{HLLC flux}\label{sec:hllc}

\begin{figure}[htbp]
  \centering
  \psfrag{O}{$0$}
  \psfrag{t}{$t$}
  \psfrag{x}{$x$}
  \psfrag{UL}{$w_L$}
  \psfrag{UR}{$w_R$}
  \psfrag{US}{$w^*$}
  \psfrag{SL}{$s_L$}
  \psfrag{SR}{$s_R$}
  \psfrag{ULS}{$w_L^*$}
  \psfrag{URS}{$w_R^*$}
  \psfrag{SS}{$s^*$}
  \includegraphics[width=5cm]{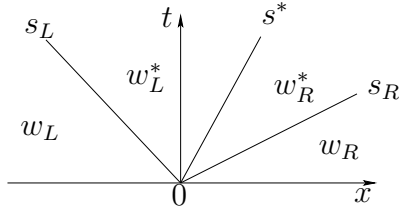}
  \caption{Approximate Riemann fan corresponding to the HLLC scheme.}
  \label{fig:HLLCfan}
\end{figure}

The HLL scheme presented briefly in the previous section was later improved by Toro, Spruce and Speares \cite{Toro1994}. Their modification concerns essentially the structure of the Riemann fan which is depicted on Figure \ref{fig:HLLCfan}. Namely, they introduced a contact discontinuity between two shock waves of the HLL scheme. That is why the novel scheme was called the HLLC scheme \cite{Fraccarollo1995}.

Here we do not provide details on the derivation of the HLLC scheme and refer to the original articles and others which can fill this gap \cite{Batten1997, Kim2007}.

We consider the same Riemann problem (\ref{eq:riemann}). In the HLLC approximation, the solution to this Riemann problem consists of three waves with speeds $s_L$, $s^*$ and $s_R$ separating four constant states $w_L$, $w_L^*$, $w_R^*$ and $w_R$. Wave speeds $s_{L, R}$ are estimated as in previous section \ref{sec:hll}, while $s^*$ is given by the formula
\begin{equation*}
  s^* = \frac{s_LH_R(u_R - s_R) - s_RH_L(u_L - s_L)}{H_R(u_R - s_R) - H_L(u_L - s_L)}.
\end{equation*}
The intermediate states $w_{L,R}^* = (H_{L,R}^*, (Hu)_{L,R}^*, (Hv)_{L,R}^*)$ are computed as follows:
\begin{equation*}
  \begin{array}{l}
        H_{L,R}^* = \frac{s_{L,R} - u_{L,R}}{s_{L,R} - s^*}H_{L,R}, \\
        (Hu)_{L,R}^* = \frac{s_{L,R} - u_{L,R}}{s_{L,R} - s^*}(Hu)_{L,R} + 
        \frac{g}{2}H_{L,R}^2\frac{(2s_{L,R}-s^*-u_{L,R})(s^*-u_{L,R})}{(s_{L,R}-s^*)^3},\\
        (Hv)_{L,R}^* = \frac{s_{L,R} - u_{L,R}}{s_{L,R} - s^*}(Hv)_{L,R} + 
        \frac{g}{2}H_{L,R}^2\frac{(2s_{L,R}-s^*-u_{L,R})(s^*-u_{L,R})}{(s_{L,R}-s^*)^3}.
  \end{array}
\end{equation*}
Finally, the numerical flux of the HLLC scheme is defined as
\begin{equation*}
  \Phi_{HLLC} (w_L, w_R) := \left\{
    \begin{array}{l}
      F_L, \quad s_L \geq 0, \\
      F_L^* := F_L + s_L(w_L^* - w_L), \quad s_L < 0 \leq s^*, \\
      F_R^* := F_R + s_R(w_R^* - w_R), \quad s^* < 0 \leq s_R, \\
      F_R, \quad s_R < 0.
    \end{array}
  \right.
\end{equation*}

\subsection{Semidiscrete scheme}

Introducing the cell averages $\w_K$ and numerical fluxes into (\ref{eq:conservlaw}) yields for the integral conservation law 
\begin{equation*}
  \od{\w_K}{t} + \sum_{L\in\N(K)} \frac{\area(L\cap K)}{\vol(K)} \Phi(\w_K, \w_L; \n_{KL}) =  \frac{1}{\vol(K)} \int_{K}\S(\w) \;d\Omega\;.
\end{equation*}
We denote by $\S_K$ the approximation of the quantity $\frac{1}{\vol(K)} \int_{K}\S(\w) \;d\Omega$. The source term discretization is discussed in Section~\ref{sec:source}. Thus, the following system of ordinary differential equations (ODE) is called a semi-discrete FV method:
\begin{equation}\label{eq:semidiscrete1}
  \od{\w_K}{t} + \sum_{L\in\N(K)} \frac{\area(L\cap K)}{\vol(K)} \Phi(\w_K, \w_L; \n_{KL}) =  \S_K, \quad \forall K\in\T\;.
\end{equation}
The initial condition for this system is given by projecting (\ref{eq:initialcond}) onto the space of piecewise constant functions
\begin{equation*}
  \w_K(0) = \frac{1}{\vol{K}} \int_{K} \w_0(x) \; d\Omega\;.
\end{equation*}
This system of ODE should also be discretized. There is a variety of explicit and implicit time integration methods. Let $\w_K^n$ denote a numerical approximation of the cell average solution in the control volume $K$ at time $t^n = n\Delta t$. The simplest time integration method is the forward Euler scheme
\begin{equation*}
  \od{\w_K}{t} \cong \frac{\w^{n+1}_K - \w^n_K}{\Delta t}\; .
\end{equation*}
When applied to (\ref{eq:semidiscrete1}) it produces the fully-discrete FV scheme: 
\begin{equation}\label{2.100}
  \frac{\w^{n+1}_K - \w^n_K}{\Delta t} + \sum_{L\in\N(K)} \frac{\area(L\cap K)}{\vol(K)} \Phi(\w_K^n, \w_L^n; \n_{KL}) =  \S_K^n, \quad \forall K\in\T\;.
\end{equation}
The time discretization used in this study is detailed in Section~\ref{sec:time}.

\subsection{Run-up algorithm}\label{sec:runup}

As already pointed out above, the NSWE are strictly hyperbolic if $H > 0$, i.e. when some water is present. The shoreline position is given by the implicit relation $H(\x,t) = 0$. At these locations the system loses its strict hyperbolicity. Finally, in dry regions, $H < 0$, the system is non-hyperbolic, i.e. ill-posed. All these facts mean that there are some major theoretical difficulties in considering the inundation problem.

Very often some ad-hoc artificial techniques are implemented to circumvent run-up and run-down problems (``slot technique'' of Madsen et al. \cite{Madsen1997}, algorithm of Hibberd-Peregrine \cite{Hibberd1979}, use of coordinate transformations \cite{Ozkan-Haller1997} and so on) -- see also \cite{MarcheBonneton2007}.

The shoreline boundary conditions have a very simple analytical form:
\begin{equation*}
  H (\x_s(t), t) = 0, \qquad \od{\x_s}{t} = \u (\x_s(t), t),
\end{equation*}
where $\x_s(t)$ is the shoreline position.

\begin{figure}
  \centering
          \psfrag{hL}{\tiny{$(H_L, u_L)$}}
          \psfrag{hR}{\tiny{$(H_R, u_R) \equiv 0$}}
          \psfrag{x}{\tiny{$x$}}
          \psfrag{xs}{\tiny{$x_s$}}
          \psfrag{x0}{\tiny{$x_{s-1}$}}
          \psfrag{us}{\tiny{$u_s$}}
        \includegraphics[width=0.79\textwidth]{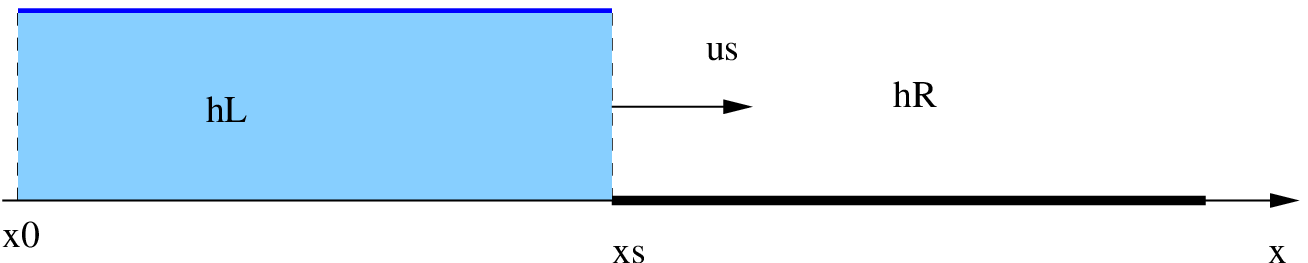}
        \includegraphics[width=0.79\textwidth]{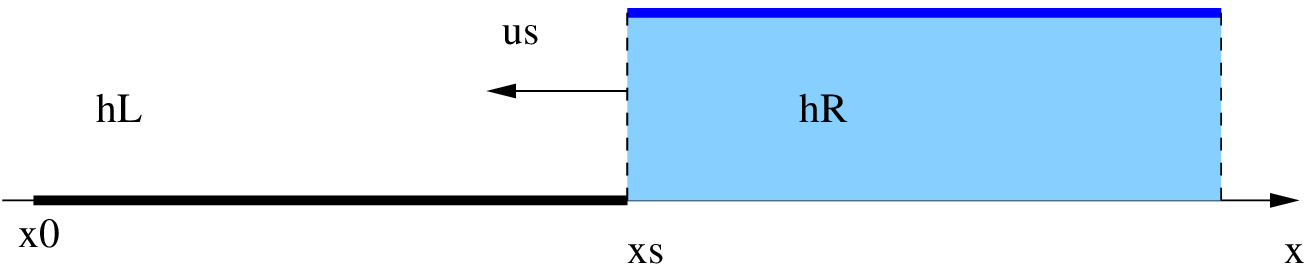}
        \caption{Shoreline left and right Riemann problem.}
        \label{fig:leftrightRiemann}
\end{figure}

\begin{figure}
  \centering
    \psfrag{xs}{\tiny{$x_s$}}
    \psfrag{uR}{\tiny{$(H_R, u_R)\equiv 0$}}
    \psfrag{uL}{\tiny{$(H_L, u_L)$}}
    \psfrag{sR}{\tiny{$s_R = u_L + 2c_L$}}
    \psfrag{sL}{\tiny{$s_L = u_L - c_L$}}
    \psfrag{x}{\tiny{$x$}}
    \psfrag{x0}{\tiny{$x_{s-1}$}}
    \psfrag{uS}{\tiny{$(H^*, u^*)$}}
    \psfrag{us}{\tiny{$u_s$}}
        \includegraphics[width=0.69\textwidth]{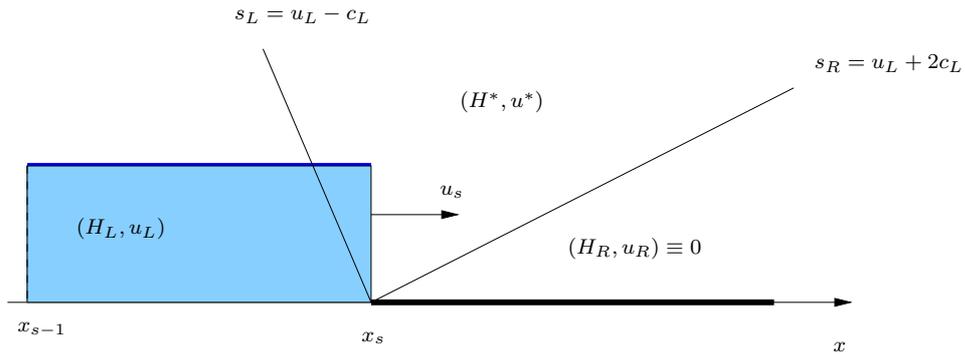}
        \caption{Shoreline Riemann problem (left) and wave propagation speeds.}
        \label{fig:shore}
\end{figure}

The algorithm proposed by Brocchini et al. \cite{Brocchini2001} was chosen for the \VOLNA code. It is based on the shoreline Riemann problem shown in Figure \ref{fig:leftrightRiemann} \cite{Stoker1957}:
\begin{equation*}
  \Rc_{\mathrm{left}}(\w_L): \left\{
    \begin{array}{l}
      \pd{\w}{t} + \pd{F(\w)}{x} = 0, \\
      \w(x,0) = \left\{
        \begin{array}{l}
          \w_L, \quad x < 0, \\
          0, \quad x > 0.
        \end{array}
      \right.
    \end{array}
  \right.
        \Rc_{\mathrm{right}}(\w_R): \left\{
    \begin{array}{l}
      \pd{\w}{t} + \pd{F(\w)}{x} = 0, \\
      \w(x,0) = \left\{
        \begin{array}{l}
          0, \quad x < 0, \\
          \w_R, \quad x > 0.
        \end{array}
      \right.
    \end{array}
  \right.
\end{equation*}
The main idea to solve the shoreline Riemann problem is to pass to the limit $\w_L\to 0$ or $\w_R\to 0$ in the solution to the classical Riemann problem (\ref{eq:riemann}). Technical details can be found in \cite{Brocchini2001}. However, we do not need to know the complete solution. It is sufficient to extract the wave propagation speeds at the shoreline (see Figure \ref{fig:shore}). These analytically determined speeds are imposed in an approximate Riemann solver when a wet/dry transition is detected.

Consider two control volumes $K$ and $L$ which share a common face. We must find the numerical flux $\Phi (\w_K, \w_L, \n_{KL})$ across this face. Let us summarize the key points of the method:
\begin{description}
  \item[\textbf{Wet/wet interface:}] If $H_L > 0$ and $H_R > 0$, we apply in the usual way an approximate Riemann solver which gives the numerical flux $\Phi$.
  
  \item[\textbf{Dry/dry interface:}] If $H_L = H_R = 0$, we just return the zero flux $\Phi = 0$ since there is no flow between two dry cells.
  
  \item[\textbf{Wet/dry interface:}] If $H_R = 0$ and $H_L > 0$, we have a situation corresponding to the left shoreline Riemann problem. Its solution yields the following choice of the wave speeds:
  \begin{equation*}
    s_L := (\u_L\cdot\n_{KL}) - c_L, \qquad
    s_R := (\u_L\cdot\n_{KL}) + 2c_L.
  \end{equation*}
  Then we apply the HLL or the HLLC scheme with the above values of $s_L$ and $s_R$ (see Figure~\ref{fig:shore} for illustration).
  
  \item[\textbf{Dry/wet interface:}] If $H_R > 0$ and $H_L = 0$, we have a situation symmetric to the previous case. One must solve the right shoreline Riemann problem. It provides the following speeds:
  \begin{equation*}
    s_L := (\u_R\cdot\n_{KL}) - 2c_R, \qquad
    s_R := (\u_R\cdot\n_{KL}) + c_R.
  \end{equation*}
  Here again, the HLL or the HLLC scheme is applied.
\end{description}

We would like to underline the simplicity of this approach. In fact, there is no special treatment for the interface. This algorithm is run uniformly in the whole computational domain leading to an easy and robust implementation. We validate this method in sections \ref{sub:catalina1} -- \ref{sub:catalina3}.

\subsection{Source terms discretization}\label{sec:source}

In this section we discuss some issues related to the source term discretization and we explain a technique to remedy them.

Source terms of the form $gH\grad h$ arise in the horizontal momentum conservation equation (\ref{eq:gov2}). Obviously, this term is equal to zero when the bottom is flat ($h = const$). However, it is not the case in real world applications. The magnitude of this term is proportional to the bed slope and may take large values when abrupt changes are present in the bathymetry.

Another profound property of NSWE is that the system (\ref{eq:gov1}), (\ref{eq:gov2}) admits non-trivial steady states. They can be determined from the following steady equations:
\begin{equation*}
  \left\{
    \begin{array}{l}
      \div(H\u) = 0, \\
      \div\bigl(H\u\otimes\u + \frac{g}{2}H^2\bigr) = gH\grad h
    \end{array}
  \right.
\end{equation*}
It is not so trivial to find analytical solutions to these equations. However, an ideal numerical scheme should preserve them. Recall that for 1D flows it is possible to describe the whole family of steady states and this information can be used to design efficient source term discretizations \cite{LeRoux1998, Vazquez-Cendon1999}. Since most applications require a 2D solver, we address this problem directly in 2D.

As explained above, it seems to be extremely difficult to construct a scheme which preserves exactly all steady state solutions. Thus, we have to simplify the problem. We will focus our attention on a simple class of steady solutions which are called in the literature ``lake at rest'':
\begin{equation}\label{eq:wellB}
  \u = 0, \qquad \eta := H - h = const.
\end{equation}
The last relations can be expressed in discrete variables:
\begin{equation}\label{eq:wellBdiscr}
  \u_K = \u_L = 0, \qquad
  H_K - h_K = H_L - h_L = const.
\end{equation}
We briefly present the method chosen for our code and developed in \cite{Audusse2004, Audusse2005}. It is based on the idea of the interface hydrostatic reconstruction.

The well-balanced algorithm takes as input the vector of conservative variables $\{ \w_K\}_{K\in\T}$, bathymetry data $\{ h_K\}_{K\in\T}$ and is composed of the following steps:
\begin{itemize}
  \item Assume that the control volumes $K$ and $L$ share a common face $K\cap L$. In this case, the interface bathymetry is defined as $h_{KL}^* := \min (h_K, h_L)$. This step is done only once at the initialization stage.
  
  \item The hydrostatic reconstructed interface water depth is given by
  \begin{equation*}
    H_{KL}^* = (H_K - h_K + h_{KL}^*)_+, \quad \mathrm{ where } \quad
    z_+ = \max (z, 0).
  \end{equation*}
  From the dicrete interpretation (\ref{eq:wellBdiscr}) of the well-balanced condition (\ref{eq:wellB}), we define a new vector of the interface conservative variables:
  \begin{equation}\label{eq:interface}
    \w_{KL}^* := \begin{pmatrix}
      H_{KL}^* \\ 
      H_{KL}^*\u_K \\
    \end{pmatrix}.
  \end{equation}
  
  \item From the balance of hydrostatic forces $\grad\bigl(\frac{g}{2}H^2\bigr) = gH\grad h$, the adapted discretization of the source terms is introduced:
  \begin{equation*}
    \S_K^* (\w_K, \w_{KL}^*, \n_{KL}) := \begin{pmatrix}
      0 \\
      \frac{g}{2}(H_{KL}^{*2} - H_K^2)\n_{KL} \\
    \end{pmatrix}
  \end{equation*}
  
  \item The well-balanced scheme is obtained by replacing cell-centered values $\w_K$ by new interface values (\ref{eq:interface}):
        \begin{multline*}
          \frac{\w^{n+1}_K - \w^n_K}{\Delta t} + \sum_{L\in\N(K)} \frac{\area(L\cap
          K)}{\vol(K)} \Phi(\w_{KL}^{*,\; n}, \w_{LK}^{*,\; n}; \n_{KL}) = \\ 
\S_K^* (\w_K^n, \w_{KL}^{*,\; n}, \n_{KL}), \quad \forall K\in\T\;.
        \end{multline*}
\end{itemize}

It can be proven \cite{Audusse2005} that the hydrostatic reconstruction strategy preserves the ``lake at rest'' solutions and ensures the positivity property. We describe here only the first-order algorithm for the sake of simplicity. The extension to second order can be found in the original papers and in \cite{Audusse2004a}.

\subsection{Time discretization}\label{sec:time}

In the previous sections we considered the spatial discretization procedure with a FV scheme. It is a common practice in solving time-dependent PDEs to first discretize the spatial variables. This approach is called method of lines:
\begin{equation}\label{eq:semidiscrete}
  w_t + \dx f(w) = S(w) \stackrel{\mbox{FV}}{\Longrightarrow} w_t = \L(w)
\end{equation}
In order to obtain a fully discrete scheme, we must discretize the time evolution operator. In the present work we chose the so-called Strong Stability-Preserving (SSP) time discretization methods described in \cite{Shu1988, Gottlieb2001, Spiteri2002}. Historically these methods were called Total Variation Diminishing (TVD) time discretizations.

The main idea behind SSP methods is to assume that the first order forward Euler method is strongly stable (see the definition below) under a certain norm for the method of lines ODE (\ref{eq:semidiscrete}). Then, we try to find a higher order scheme. Usually the relevant norm is the total variation\footnote{The notion of total variation is used essentially for 1D discrete solutions.} norm:
\begin{equation*}
  \TV(w^n) := \sum_{j} \abs{w_j^n - w_{j-1}^n}
\end{equation*}
and TVD discretizations have the property $\TV(w^{n+1}) \leq \TV(w^n)$.

\begin{remark}
  Special approaches are needed for hyperbolic PDEs since they contain discontinuous solutions and the usual linear stability analysis is inadequate. Thus a stronger measure of stability is usually required:
  \begin{definition}
    A sequence $\set{w^n}$ is said to be \textbf{strongly stable} in a given norm $\norm{\cdot}$ provided that $\norm{w^{n+1}} \leq \norm{w^n}$ for all $n\geq 0$.
  \end{definition}
\end{remark}

A general $m$-stage Runge-Kutta method for (\ref{eq:semidiscrete}) can be written in the form
\begin{eqnarray}\label{eq:rungekutta}
  w^{(0)} &=& w^n, \\
  w^{(i)} &=& \sum_{k=0}^{i-1} \Bigl(\alpha_{i,k}w^{(k)}+\Delta t\beta_{i,k}\L(w^{(k)})\Bigr),
  \quad \alpha_{i,k}\geq 0, \quad i=1,\ldots,m, \\
  w^{n+1} &=& w^{(m)}.\label{eq:rungelast}
\end{eqnarray}

In \cite{Shu1988a} the following result is proved
\begin{theorem}
  If the forward Euler method is strongly stable under the CFL restriction $\Delta t\leq\Delta t_{FE}$
  \begin{equation*}
    \norm{w^n + \Delta t\L(w^n)} \leq \norm{w^n},
  \end{equation*}
  then the Runge-Kutta method (\ref{eq:rungekutta}) -- (\ref{eq:rungelast}) with $\beta_{i,k}\geq 0$ is SSP, $\norm{w^{n+1}}\leq\norm{w^n}$, provided the following CFL restriction is fulfilled:
  \begin{equation*}
    \Delta t \leq c \Delta t_{FE}, \quad
    c = \min_{i,k} \frac{\alpha_{i,k}}{\beta_{i,k}}.
  \end{equation*}
\end{theorem}

Here we give a few examples of SSP schemes which are commonly used in applications (optimality is in the sense of CFL condition):
\begin{itemize}
\item Optimal second order two-stage SSP-RK(2,2) scheme with CFL $= 1$:
  \begin{eqnarray*}
    w^{(1)} &=& w^{(n)} + \Delta t \L(w^{(n)}), \\
    w^{(n+1)} &=& \frac12 w^{(n)} + \frac12 w^{(1)} + \frac12\Delta t \L(w^{(1)});
  \end{eqnarray*}
\item Optimal third order three-stage SSP-RK(3,3) scheme with CFL $= 1$:
  \begin{eqnarray*}
    w^{(1)} &=& w^{(n)} + \Delta t \L(w^{(n)}), \\
    w^{(2)} &=& \frac34 w^{(n)} + \frac14 w^{(1)} + \frac14\Delta t \L(w^{(1)}), \\
    w^{(n+1)} &=& \frac13 w^{(n)} + \frac23 w^{(2)} + \frac23\Delta t \L(w^{(2)});
  \end{eqnarray*}
\item Third order four-stage SSP-RK(3,4) scheme with CFL $= 2$:
  \begin{eqnarray*}
    w^{(1)} &=& w^{(n)} + \frac12\Delta t \L(w^{(n)}), \\
    w^{(2)} &=& w^{(1)} + \frac12\Delta t \L(w^{(1)}), \\
    w^{(3)} &=& \frac23w^{(n)} + \frac13w^{(2)} + \frac16\Delta t \L(w^{(n)}), \\
    w^{(n+1)} &=& w^{(3)} + \frac12\Delta t \L(w^{(3)}).
  \end{eqnarray*}
\end{itemize}
The linear absolute stability region for the RK and SSP-RK schemes is the same. However the nonlinear absolute stability regions are quite different \cite{Cartwright1992}.

We tested these different schemes in our numerical code and decided to adopt SSP-RK(3,4) due to its accuracy and wide stability region. In our opinion this scheme represents a very good trade-off between precision and robustness.

\subsection{Second order extension}

If we analyze the above scheme, we understand that in fact, we have only one degree of freedom per data storage location. Hence, it seems that we can expect to be first order accurate at most. In the numerical community first order schemes are generally considered to be too inaccurate for most quantitative calculations. Of course, we can always make the mesh spacing extremely small but it cannot be a solution since it makes the scheme inefficient. From the theoretical point of view the situation is even worse since an $\O(h^{\frac12})$ $L_1$-norm error bound for the monotone and E-flux schemes \cite{Osher1984} is known to be sharp \cite{Peterson1991}, although an $\O(h)$ solution error is routinely observed in numerical experiments. On the other hand, Godunov has shown \cite{Godunov1959} that all linear schemes that preserve solution monotonicity are at most first order accurate. This rather negative result suggests that a higher order accurate scheme has to be essentially nonlinear in order to attain simultaneously a monotone resolution of discontinuities and high order accuracy in continuous regions.

A significant breakthrough in the generalization of FV methods to higher order accuracy is due to N.E. Kolgan \cite{Kolgan1972, Kolgan1975} and van Leer \cite{Leer1979}. They proposed a kind of post-treatment procedure currently known as solution \textit{reconstruction} or MUSCL (Monotone Upstream-centered Scheme for Conservation Laws) scheme. In the above papers the authors used linear reconstruction (it will be chosen in this study as well) but this method has already been extended to quadratic approximations in each cell \cite{Barth1990}.


\subsubsection{Historical remark}

In general, authors of numerical articles which use the MUSCL scheme often cite the paper by van Leer \cite{Leer1979}. It is commonly believed in the scientific community that B. van Leer was first to propose the gradient reconstruction and slope limiting ideas. Because of unfortunate political reasons, N.E. Kolgan's work \cite{Kolgan1972, Kolgan1975} remained unknown for a long time. We would like to underline the fact that the first publication of Kolgan came out seven years before van Leer's paper. Van Leer seems to be aware of this situation since in his recent review paper \cite{Leer2006} one can find ``A historical injustice'' section:
\begin{quote}
  ``It has been pointed out to me by Dr. Vladimir Sabelnikov, formerly of TsAGI, the Central Aerodynamical National Laboratory near Moscow, that a scheme closely resembling MUSCL (including limiting) was developed in this laboratory by V. P. Kolgan (1972). Kolgan died young; his work apparently received little notice outside TsAGI.''
\end{quote}


\subsubsection{TVD and MUSCL schemes}

There is a property of scalar nonlinear conservation laws, which was probably observed for the first time by P.~Lax \cite{Lax1973}:
{\it The total increasing and decreasing variations of a differentiable solution between any pair of characteristics are conserved.}
In the presence of shock waves, information is lost and the total variation decreases. For compactly supported or periodic solutions, one can establish the following inequality
\begin{equation}\label{eq:tvdcont}
  \int\limits_{-\infty}^{+\infty}\abs{dw(x,t_2)} \leq \int\limits_{-\infty}^{+\infty}\abs{dw(x,t_1)}, \quad
  t_2 \geq t_1.
\end{equation}
This motivated Harten \cite{Harten1983} to introduce the notion of discrete total variation of numerical solution $w_h := \set{w_j}$
\begin{equation*}
  TV(w_h) := \sum_{j} \abs{w_{j+1} - w_j},
\end{equation*}
and the discrete counterpart to (\ref{eq:tvdcont})
\begin{equation*}
  TV(w_h^{n+1}) \leq TV(w_h^n).
\end{equation*}
If this property is fulfilled, then a FV scheme is said to be total variation diminishing (TVD). The following theorem was proved in \cite{Harten1983}:
\begin{theorem}
  (i) Monotone schemes are TVD; (ii) TVD schemes are monotonicity preserving, i.e. the number of solution extrema is preserved in time.
\end{theorem}

\begin{remark}
  From the mathematical point of view it would be more correct to say ``the total variation non-increasing (TVNI) scheme'' but the ``wrong'' term TVD is generally accepted in the scientific literature.
\end{remark}

In one space dimension the construction of TVD schemes is not a problem anymore. Let us recall that in this study we are rather interested in two space dimensions (or even three in future work). In these cases the situation is considerably more complicated. Even if we consider the simplest case of structured cartesian meshes and apply a 1D TVD scheme on a dimension-by-dimension basis, a result of Goodman and Leveque shows \cite{Goodman1985} that TVD schemes in two or more space dimensions are only first order accurate. Motivated by this negative result, weaker conditions yielding solution monotonicity preservation should be developed.

In this article we describe the construction and practical implementation of a second-order nonlinear scheme on unstructured (possibly highly distorted) meshes. The main idea is to find our solution as a piecewise affine function on each cell. This kind of linear reconstruction operators on simplicial control volumes often exploit the fact that the cell average is also a pointwise value of any valid (conservative) linear reconstruction evaluated at the center of a simplex. This reduces the reconstruction problem to that of gradient estimation given cell averaged data. In this case, we express the reconstruction in the form
\begin{equation}\label{eq:reconstruct}
  \w_K (\x) = \bar\w_K + (\grad\w)_K\cdot(\x - \x_0), \quad K\in\T \;,
\end{equation}
where $\bar\w_K$ is the cell averaged value given by the FV method, $(\grad\w)_K$ is the solution gradient estimate (to be determined) on the cell $K$, $\x\in K$ and the point $\x_0$ is chosen to be the center for the simplex $K$.

It is important to note that with this type of representation (\ref{eq:reconstruct}) we remain absolutely conservative, i.e.
\begin{equation*}
  \frac{1}{\vol(K)}\int_{K} \w_K(\x)\;d\Omega \equiv \bar\w_K
\end{equation*}
due to the choice of the point $\x_0$. This point is crucial for FVs because of intrinsic conservative properties of this method.

In the next sections we describe briefly two common techniques: Green-Gauss integration and least squares methods for solution gradient estimation on each cell. There are other available techniques. We can mention here an implicit gradient reconstruction method proposed in \cite{Musaferija1996} and reused later in \cite{Archambeau2004}. We decided not to implement this approach in our research code since this procedure is computationally expensive\footnote{In order to reconstruct the solution gradient we have to solve a linear system of equations. Recall that the gradient is estimated at each time step on each control volume. This factor slows down considerably explicit time discretizations.}.

\subsubsection{Green-Gauss gradient reconstruction}\label{sec:ggo}

\begin{figure}[htbp]
  \centering
  \psfrag{O}{$O$}
  \psfrag{K}{$K$}
  \psfrag{M}{$\partial K$}
  \psfrag{n}{$\n$}
  \psfrag{e}{$e$}
  \psfrag{N}{$N_1$}
  \psfrag{P}{$N_2$}
  \includegraphics[width=5cm]{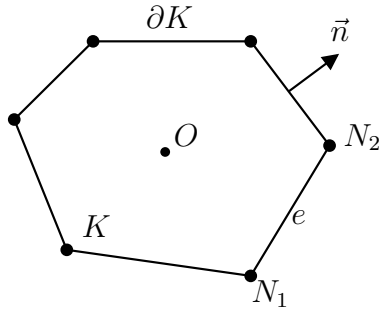}
  \caption[Illustration for Green-Gauss gradient reconstruction]{Illustration for Green-Gauss gradient reconstruction. Control volume $K$ with barycenter $O$ and exterior normal $\n$.}
  \label{fig:greengauss}
\end{figure}

This gradient reconstruction technique can be easily implemented on simplicial meshes. It is based on two very simple ideas: the mean value approximation and Green-Gauss-Ostrogradsky formula.

Consider a control volume $K$ with barycenter $O$. The exterior normal to an edge $e\in\partial K$ is denoted by $\n_e$. This configuration is depicted on \figurename~\ref{fig:greengauss}. In order to estimate the solution gradient on $K$ (or in other words, to estimate its value at the center $O$) we make the following mean value approximation
\begin{equation*}
  (\grad\w)_K = \left.(\grad\w)\right|_{O} \cong \frac{1}{\vol(K)} \int_K \grad\w\; d\Omega,
\end{equation*}
and apply Green-Gauss-Ostrogradsky formula
\begin{multline*}
  (\grad\w)_K \cong \frac{1}{\vol(K)} \int_{\partial K} \w\otimes\n\; d\sigma =
  \frac{1}{\vol(K)}\sum_{e\in\partial K} \int_e \w\otimes\n_e\; d\sigma \cong \\
  \sum_{e\in\partial K}\frac{\area(e)}{\vol(K)} \left.\w\right|_{e/2}\otimes\n_e,
\end{multline*}
where $\left.\w\right|_{e/2}$ denote the solution value at the face (or edge in 2D) centroid. The face value needed to compute the
reconstruction gradient can be obtained from a weighted average of the values at the vertices on the face \cite{Holmes1989}. In 2D it simply becomes
\begin{equation*}
  \left.\w\right|_{e/2} = \frac{\w_{N_1} + \w_{N_2}}{2}.
\end{equation*}
This approximation yields the following formula for gradient estimation:
\begin{equation*}
  (\grad\w)_K \cong \sum_{e\in\partial K}\frac{\area(e)}{\vol(K)} \frac{(\w_{N_1} + \w_{N_2})}{2}\otimes\n_e.
\end{equation*}
The gradient calculation is exact whenever the numerical solution varies linearly over the support of the reconstruction.

This procedure requires the knowledge of the solution values at the mesh nodes $\set{N_i}$. Since a cell centered FV scheme provides data located at cell centers, an interpolation technique is needed. The quality of Green-Gauss gradient reconstruction greatly depends on the chosen interpolation method. The method chosen here is explained in Section~\ref{sec:interpolation}.

\subsubsection{Least-squares gradient reconstruction method}\label{sec:lsq}

\begin{figure}[htbp]
  \centering
  \psfrag{K}{$K$}
  \psfrag{T}{$T_1$}
  \psfrag{S}{$T_2$}
  \psfrag{U}{$T_3$}
  \psfrag{O}{$O$}
  \psfrag{O1}{$O_1$}
  \psfrag{O2}{$O_2$}
  \psfrag{O3}{$O_3$}
  \includegraphics[width=5cm]{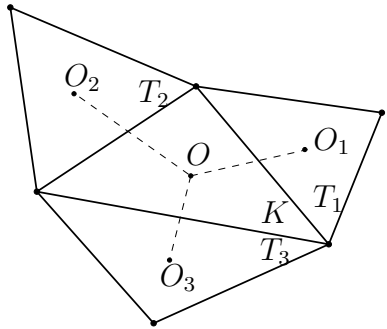}
  \caption[Illustration for least-squares gradient reconstruction.]{Illustration for least-squares gradient reconstruction. A triangle control volume with three adjacent neighbors is depicted.}
  \label{fig:leastsq}
\end{figure}

In this section we consider a triangle\footnote{Generalization to other simplicial control volumes is straightforward.} control volume $K$ with three adjacent neighbors $T_1$, $T_2$ and $T_3$. Their barycenters are denoted by $O(\x_0)$, $O_1(\x_1)$, $O_2(\x_2)$ and $O_3(\x_3)$ respectively. In the following we denote by $\w_i$ the solution value at the centers $O_i$:
\begin{equation*}
  \w_i := \w(\x_i), \quad \w_0 := \w(\x_0).
\end{equation*}

Our purpose here is to estimate $\grad\w = (\partial_x\w, \partial_y\w)$ on the cell $K$. Using Taylor formula, we can write down the three following relations:
\begin{eqnarray}\label{eq:firstconstraint}
  \w_i - \w_0 &=& (\grad\w)_K\cdot(\x_i - \x_0) + \O(h^2), \qquad i=1,2,3. 
\end{eqnarray}
If we drop higher order terms $\O(h^2)$, these relations can be viewed as a linear system of three equations for two unknowns\footnote{This simple estimation is done for the scalar case only $\w = (w)$. For more general vector problems the numbers of equations and unknowns must be changed depending on the dimension of vector $\w$.} $(\partial_x\w, \partial_y\w)$.  This situation is due to the fact that the number of edges incident to a simplex mesh in $\R^d$ is greater or equal (in this case see Remark~\ref{rem:square}) to $d$ thereby producing linear constraint equations (\ref{eq:firstconstraint}) which will be solved analytically here in a least squares sense.

First of all, each constraint (\ref{eq:firstconstraint}) is multiplied by a weight $\omega_i\in(0,1)$ which will be chosen below to account for distorted meshes. In matrix form our non-square system becomes
\begin{equation*}
  \begin{pmatrix}
    \omega_1\Delta x_1 & \omega_1\Delta y_1 \\
    \omega_2\Delta x_2 & \omega_2\Delta y_2 \\
    \omega_3\Delta x_3 & \omega_3\Delta y_3 \\
  \end{pmatrix}
  (\grad\w)_K =
  \begin{pmatrix}
    \omega_1 (\w_1 - \w_0) \\
    \omega_2 (\w_2 - \w_0) \\
    \omega_3 (\w_3 - \w_0) \\
  \end{pmatrix},
\end{equation*}
where $\Delta x_i = x_i - x_0$, $\Delta y_i = y_i - y_0$.
For further developments it is convenient to rewrite our constraints in abstract form
\begin{equation}\label{eq:abstract}
  [\vec{L_1},\; \vec{L_2}]\cdot (\grad\w)_K = \vec{f}.
\end{equation}
We use a normal equation technique in order to solve symbolically this abstract form in a least squares sense. Multiplying on the left both sides of (\ref{eq:abstract}) by $[\vec{L_1} \vec{L_2}]^t$ yields
\begin{equation}\label{eq:squaresystem}
  G(\grad\w)_K = \vec{b}, \quad
  G = (l_{ij})_{1\leq i,j\leq 2} =
  \begin{pmatrix}
    (\vec{L_1}\cdot\vec{L_1}) & (\vec{L_1}\cdot\vec{L_2}) \\
    (\vec{L_2}\cdot\vec{L_1}) & (\vec{L_2}\cdot\vec{L_2}) \\
  \end{pmatrix}
\end{equation}
where $G$ is the Gram matrix of vectors $\set{\vec{L_1},\vec{L_2}}$ and
$
\vec{b} =
\begin{pmatrix}
  (\vec{L_1}\cdot\vec{f}) \\
  (\vec{L_2}\cdot\vec{f}) \\
\end{pmatrix}.
$
The so-called normal equation (\ref{eq:squaresystem}) is easily solved by Cramer's rule to give the following result
\begin{equation*}
  (\grad\w)_K = \frac{1}{l_{11}l_{22} - l_{12}^2}
  \begin{pmatrix}
    l_{22}(\vec{L_1}\cdot\vec{f}) - l_{12}(\vec{L_2}\cdot\vec{f}) \\
    l_{11}(\vec{L_2}\cdot\vec{f}) - l_{12}(\vec{L_1}\cdot\vec{f}) \\
  \end{pmatrix}.
\end{equation*}
The form of this solution suggests that the least squares linear reconstruction can be efficiently computed without the need for storing a non-square matrix.

Now we discuss the choice of weight coefficients $\set{\omega_i}_{i=1}^3$. The basic idea is to attribute bigger weights to cells barycenters closer to the node $N$ under consideration. One of the possible choices consists in taking a harmonic mean of respective distances $r_i = ||\x_i - \x_N||$. This purely metric argument takes the following mathematical form:
\begin{equation*}
  \omega_i = \frac{||\x_i - \x_N||^{-k}}{\sum_{j=1}^3||\x_j - \x_N||^{-k}},
\end{equation*}
where $k$ in practice is taken to be one or two (in our code we choose $k=1$).

\begin{remark}\label{rem:square}
  When a triangle shares an edge with the boundary $\partial\Omega$ (see \figurename~\ref{fig:boundcondInt} for illustration), the gradient reconstruction procedure becomes even simpler, since the number of constraints is equal to $d$ and the linear system (\ref{eq:firstconstraint}) becomes completely determined:
  \begin{eqnarray*}
    \w_i - \w_0 &=& (\grad\w)_K\cdot(\x_i - \x_0) + \O(h^2), \qquad i=1,2. 
  \end{eqnarray*}
  In component form it reads
  \begin{equation*}
    \begin{pmatrix}
      x_1 - x_0 & y_1 - y_0 \\
      x_2 - x_0 & y_2 - y_0 \\
    \end{pmatrix}
    (\grad\w)_K =
    \begin{pmatrix}
      \w_1 - \w_0 \\
      \w_2 - \w_0 \\
    \end{pmatrix}.
  \end{equation*}
  The unique solution to this linear system is given again by Cramer's rule
  \begin{equation*}
    (\grad\w)_K = \frac{\begin{pmatrix}
        (y_2-y_0)(\w_1-\w_0) - (y_1-y_0)(\w_2-\w_0) \\
        (x_1-x_0)(\w_2-\w_0) - (x_2-x_0)(\w_1-\w_0) \\
      \end{pmatrix}}%
    {(x_1-x_0)(y_2-y_0) - (x_2-x_0)(y_1-y_0)}.
  \end{equation*}
\end{remark}

\subsubsection{Slope limiter}

The idea of incorporating limiter functions to obtain non-oscillatory resolution of discontinuities and steep gradients goes back to Boris and Book \cite{Boris1973}. When the limiter is identically equal to $1$, we have the unlimited form of the linear interpolation. In the 1D case one can easily find in the literature about 15 different limiter functions such as CHARM, minmod, superbee, van Albada and many others. On unstructured meshes the situation is quite different. In the present study we decided to choose the Barth-Jespersen limiter proposed in \cite{Barth1989}. Here we do not discuss its construction and properties but just give the final formula. We need to introduce the following notation
\begin{equation*}
  \w_K^{min} := \min_{L\in \N(K)} \w_L, \quad
  \w_K^{max} := \max_{L\in \N(K)} \w_L\; .
\end{equation*}

The limited version of (\ref{eq:reconstruct}) is given by the following modified reconstruction operator
\begin{equation*}
  \w_K (\x) = \bar\w_K + \alpha_K(\grad\w)_K\cdot(\x - \x_0), \quad K\in\T \;,
\end{equation*}
where it is assumed that $\alpha_K\in[0,1]$. Obviously, the choice $\alpha_K = 0$ corresponds to the first order scheme while $\alpha_K = 1$ is the unlimited form. Barth and Jespersen \cite{Barth1989} propose the following choice of $\alpha_K$:
\begin{equation*}
  \alpha_K^{BJ} := \min_{\forall f\in\partial K}\left\{%
    \begin{array}{lc}
      \frac{\w_K^{max} - \bar\w_K}{\w_K(\x_f) - \bar\w_K} & \mbox{if } \w_K(\x_f) > \w_K^{max}, \\
      \frac{\w_K^{min} - \bar\w_K}{\w_K(\x_f) - \bar\w_K} & \mbox{if } \w_K(\x_f) < \w_K^{min}, \\
      1 & \mbox{otherwise},
    \end{array}
  \right.
\end{equation*}
where $\x_f$ denotes the face $f$ centroid.

Although this limiter function does not fulfill all the requirements of FV maximum principle on unstructured meshes \cite{Barth2004}, it can be shown that it yields FV schemes possessing a global extremum diminishing property. Also this limiter produces the least amount of slope reduction which can be advantageous for accuracy. Note that in practice minor modifications are required to prevent near zero division for almost constant solution data.

\subsubsection{Solution interpolation to mesh nodes}\label{sec:interpolation}

We have seen above that several gradient reconstruction procedures (in particular gradient estimation on the faces) require the knowledge of the solution at mesh nodes (or vertices). This information is not directly given by the FV method since we chose the cell-centered approach.

\begin{figure}[htbp]
  \centering
  \psfrag{O}{$O_{i+1}$}
  \psfrag{P}{$O_i$}
  \psfrag{Q}{$O_{i-1}$}
  \psfrag{N}{$N$}
  \includegraphics[width=5cm]{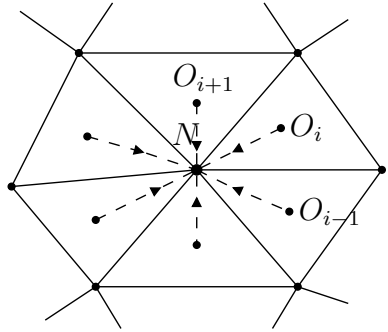}
  \caption[Triangles sharing the same vertex $N$.]{Triangles with their barycenters $O_i$ sharing the same vertex $N$.}
  \label{fig:interpnodes}
\end{figure}

Let us consider a node $N(x_n, y_n)$ of the tesselation $\T$ and a control volume $K_i$ with barycenter $O_i(x_i, y_i)$ having this node as a vertex (see \figurename~\ref{fig:interpnodes} for illustration). The MUSCL procedure provides a solution gradient on each cell. Thus, using the Taylor formula or, equivalently, the representation (\ref{eq:reconstruct}) we can estimate the solution value at the node $N$
\begin{equation}\label{eq:atnode}
  \w_N = \bar\w_{K_i} + (\grad\w)_{K_i}\cdot(\x_N - \x_i).
\end{equation}
The problem is that we will have $d(N)$ different values of the solution in the same point depending on the control volume under consideration. Here $d(N)$ is the degree of vertex $N$ in the sense of graph theory. One of the possible ways to overcome this contradiction is averaging. One interesting technique was proposed in \cite{Holmes1989}, further improved in \cite{Kim2003} and slightly modified by us. The algorithm implemented in our code is briefly described here.

First of all, let us look for the vertex value $\bar\w_N$ as a weighted sum of the values $\w_{N_i}$ computed by formula (\ref{eq:atnode}) from each surrounding cell
\begin{equation*}
  \bar\w_N = \frac{\sum_{i=1}^{d(N)}\omega_i\w_{N_i}}{\sum_{i=1}^{d(N)}\omega_i}\;.
\end{equation*}

The weighting factors $\set{\omega_i}_{i=1}^{d(N)}$ are made to satisfy the condition of zero pseudo-Laplacian
\begin{equation}\label{eq:constraint}
  L(x_n) \equiv \sum_{i=1}^{d(N)} \omega_i (x_i - x_n), \quad
  L(y_n) \equiv \sum_{i=1}^{d(N)} \omega_i (y_i - y_n)\; .
\end{equation}
These conditions have a very simple interpretation. They are imposed so that the method be exact for affine data over the stencil.

As in the original formulation by Holmes and Connell \cite{Holmes1989}, the weighting factor $\omega_i$ is written as
\begin{equation*}
  \omega_i = 1 + \Delta\omega_i\;.
\end{equation*}

The weights $\set{\omega_i}$ are determined by solving an optimization problem in which the cost-function to be minimized is defined as
\begin{equation}\label{eq:costfunc}
  \frac12\sum_{i=1}^{d(N)} \bigl(r_i\Delta\omega_i\bigr)^2 \rightarrow\min
\end{equation}
with two constraints given by (\ref{eq:constraint}). It should be noted that the cost function is slightly different from the original formulation. The difference lies in the factor of
$$
r_i^2 \equiv ||\vec{ON} - \vec{OO_i}||^2
$$
which was introduced in \cite{Kim2003}. This modification effectively allows larger values of weight $\Delta\omega_i$ for those cells closer to the node in question.

Employing the method of Lagrange multipliers, the original optimization problem, which was to minimize the cost function given by (\ref{eq:costfunc}) with the constraints (\ref{eq:constraint}), is equivalent to minimizing the function $\L$ defined by
\begin{equation*}
  \L = \frac12\sum_{i=1}^{d(N)} \bigl(r_i\Delta\omega_i\bigr)^2 - \lambda\sum_{i=1}^{d(N)} \omega_i (x_i - x_n)
  - \mu \sum_{i=1}^{d(N)} \omega_i (y_i - y_n) \rightarrow\min
\end{equation*}
which leads to
\begin{equation*}
  \Delta\omega_i = \frac{\lambda (x_i - x_n) + \mu (y_i - y_n)}{r_i^2}\;.
\end{equation*}
The two Lagrangian multipliers, $\lambda$ and $\mu$, are obtained from
\begin{equation*}
  \lambda = \frac{r_y I_{xy} - r_x I_{yy}}{I_{xx}I_{yy} - I_{xy}^2}, \quad
  \mu = \frac{r_x I_{xy} - r_y I_{xx}}{I_{xx}I_{yy} - I_{xy}^2},
\end{equation*}
where
\begin{equation*}
  r_x = \sum_{i=1}^{d(N)} (x_i - x_n), \quad
  r_y = \sum_{i=1}^{d(N)} (y_i - y_n).
\end{equation*}
\begin{equation*}
  I_{xx} = \sum_{i=1}^{d(N)} \frac{(x_i - x_n)^2}{r_i^2}, \quad
  I_{yy} = \sum_{i=1}^{d(N)} \frac{(y_i - y_n)^2}{r_i^2}, \quad
  I_{xy} = \sum_{i=1}^{d(N)} \frac{(x_i - x_n)(y_i - y_n)}{r_i^2}\;.
\end{equation*}

The last step consists in renormalizing the weights $\set{\omega_i}_{i=1}^{d(N)}$ to the range $[0,1]$.

\begin{remark}
The above algorithm is not computationally expensive since the weights $\set{\omega_i}_{i=1}^{d(N)}$ only depend on the tesselation $\T$ geometry. It means that they can be computed and stored before the main loop in time and reused during later computations.
\end{remark}

\subsection{Implementation of boundary conditions}\label{sec:bound}

\begin{figure}[htbp]
  \centering
  \psfrag{K}{$K$}
  \psfrag{Q}{$\partial\Omega$}
  \includegraphics[width=5cm]{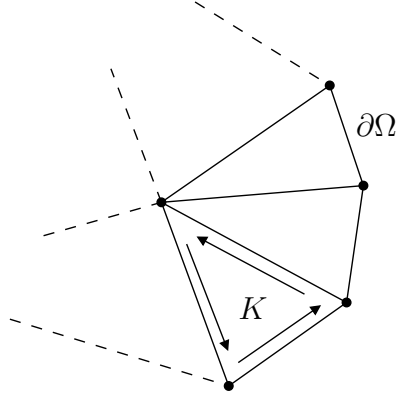}
  \caption[Control volume sharing a face with boundary $\partial\Omega$.]{Control volume sharing a face with boundary $\partial\Omega$.}
  \label{fig:boundcondInt}
\end{figure}

So far we have not discussed the implementation of boundary conditions. The flavor of the treatment of boundary conditions for hyperbolic systems is given here and we refer to \cite{Ghidaglia2005} for a general discussion. This is a very important topic since they actually determine the solution. Let us consider the space discretization of the system (\ref{1.24}) by a cell centered FV method. For instance for the time explicit discretization we have the scheme (\ref{2.100}). Of course this formula is not valid when $K$ meets the boundary of $\Om$ (see \figurename \ref{fig:boundcondInt} for illustration). When this occurs, we must find the numerical flux $\Phi(v^n_K,K,\partial\Om)$. In practice, this flux is not given by the physical boundary conditions and moreover, in general, (\ref{1.24}) is an ill-posed problem if we try to impose either $w$ or $F(w)\cdot\n$ on $\partial\Om$. This can be understood in a simple way by using the following linearization of this system:
\begin{equation}\label{bc2} 
  \frac{\partial w}{\partial t} + {\un A_n}
  \frac{\partial w}{\partial n} = 0\,, 
\end{equation}
where $\n$ represents the direction of the external normal on $K\cap\partial\Om$, ${\un A_n}$ is the advection matrix:
\begin{equation}\label{bc3}
        {\un A_n} \equiv \frac{\bigl(\partial F(w)\cdot \n\bigr)}
        {\partial w}\vert_{ w=\uvl},
\end{equation}
and $\uvl$ is the state around which the linearization is performed. When (\ref{1.24}) is hyperbolic, the matrix ${\un A_n}$ is diagonalizable on $\rit$ and by a change of coordinates, this system becomes an uncoupled set of $m$ advection equations:
\begin{equation}\label{bc4}
        \frac{\partial \xi_k}{\partial t} + \lambda_k
        \frac{\partial \xi_k}{\partial n} = 0
        \,,\quad k=1,\ldots,m\,.
\end{equation}
Here the $\lambda_k$ are the eigenvalues of ${\un A_n}$ and according to their sign, waves are going either into the domain $\Om$ ($\lambda_k < 0$) or out of the domain $\Om$ ($\lambda_k > 0$). Hence we expect that it is only possible to impose $p$ conditions on $K\cap\partial\Om$ where $p\equiv \sharp\{k\in \{1,\ldots,m\} \mbox{ such that } \lambda_k <0\}$.\\

Let us consider now a control volume $K$ which meets the boundary $\partial\Om$. We take $\uvl=w^n_K$ and write the previous linearization. We denote by $x$ the coordinate along the outer normal so that (\ref{bc2}) reads:

\begin{equation}\label{bc5} 
        \frac{\partial w}{\partial t} + {\un A_n}
        \frac{\partial w}{\partial x} = 0\,, 
\end{equation}
which happens to be the linearization of the 1D ({\it i.e.} when $nd=1$) system. First we label the eigenvalues $\lambda_k(\uvl)$ of ${\un A_n}$ by increasing order:
\begin{equation}\label{bc6}
        \lambda_1(\uvl)\leq \lambda_2(\uvl) \leq\ldots\leq \lambda_p(\uvl) < 0\leq
        \lambda_{p+1}(\uvl)\ldots\leq \lambda_m(\uvl)\,.
\end{equation}

\begin{itemize}

\item[(i)] The case $p=0$. In this case information comes from inside $\Om$ and therefore we take:
\begin{equation}\label{bc7}
        \Phi(w^n_K,K,\partial\Om) = F(w^n_K)\cdot\n_K\,.
\end{equation}
In the \cfd~(CFD) literature this is known as the ``supersonic outflow'' case.

\item[(ii)] The case $p=m$. In this case information comes from outside $\Om$ and therefore we take:
\begin{equation}\label{bc8}
        \Phi(w^n_K,K,\partial\Om) = \Phi_{given}\,,
\end{equation}
where $\Phi_{given}$ are the given physical boundary conditions. In the CFD literature this is known as the ``supersonic inflow'' case.

\item[(iii)] The case $1\leq p\leq m-1$. As already discussed, we need $p$ scalar information coming from outside of $\Om$. Hence we assume that we have on physical ground $p$ relations on the boundary:
\begin{equation}\label{bc9}
        g_l(w)=0\,,\quad l=1,\ldots,p.
\end{equation}

\end{itemize}

\begin{remark} 
The notation $g_l(w)=0$ means that we have a relation between the components of $w$. However, in general, the function $g_l$ is not given explicitly in terms of $w$. For example $g_l(w)$ could be the pressure which is not, in general, one of the components of $w$.
\end{remark}

\begin{itemize}
\item[] Since we have to determine the $m$ components of $\Phi(w^n_K,K,\partial\Om)$, we need $m-p$ supplementary scalar conditions. Let us write them as
\begin{equation}\label{bc10}
        h_l(w)=0\,,\quad l=p+1,\ldots,m.
\end{equation}
In general (\ref{bc9}) are referred to as ``physical boundary conditions'' while (\ref{bc10}) are referred to as ``numerical boundary conditions''.

Then we take:
\begin{equation}\label{bc11}
        \Phi(w^n_K,K,\partial\Om) = F(w)\cdot\n_K\,,
\end{equation}
where $w$ is solution to (\ref{bc9})-(\ref{bc10}) (see however Remark~\ref{pratique} and (\ref{bc17})).
\end{itemize}

\begin{remark} 
The system (\ref{bc9})-(\ref{bc10}) for the $m$ unknowns $w\in G$ is a $m\times m$ nonlinear system of equations. Its solvability is given by Theorem \ref{transv}.
\end{remark}

Let us first discuss the numerical boundary conditions (\ref{bc10}). By analogy with what we did on an interface between two control volumes $K$ and $L$, we take (recall that $\uvl = w^n_K)$:
\begin{equation}\label{bc12}
        \tilde l_k(\uvl)\cdot(F(w)\cdot \n_K) = \tilde l_k(\uvl)\cdot(F(w^n_K)\cdot \n_K)\,,
        \quad k=p+1,\ldots,m.
\end{equation}
In other words, we set $h_k(w)\equiv \tilde l_k(w^n_K)\cdot(F(w)\cdot \n_K)-\tilde l_k(w^n_K)\cdot(F(w^n_K)\cdot \n_K)$. We have denoted by $(\tilde l_1(\uvl),\ldots ,\tilde l_m(\uvl))$ a set of left eigenvectors of ${\un {\tilde A}_n}$: $^t{\un {\tilde A}_n}l_k(\uvl) = \lambda_k l_k(\uvl)$ and by $(r_1(\uvl),\ldots ,r_m(\uvl))$ a set of right eigenvectors of ${\un {\tilde A}_n}$: ${\un {\tilde A}_n}r_k(\uvl) = \lambda_k r_k(\uvl)$. Moreover the following normalization is taken: $\tilde l_k(\uvl)\cdot \tilde r_p(\uvl) = \delta_{k,p}$.

According to \cite{Ghidaglia2005} we have the following result on the solvability of (\ref{bc9})-(\ref{bc10}).
\begin{theorem}\label{transv}
        In the case $1\leq p\leq m-1$, assume that $\lambda_{p+1}(\uvl)> 0$, and
        \begin{equation}\label{bc13} 
                \det_{1\leq k,l\leq p}\left(\sum_{i=1}^m
                r_k^i(\uvl)\frac{\partial g_l}{\partial w_i}(\uvl)\right)\neq 0\,.
        \end{equation}
With the choice (\ref{bc12}) the nonlinear system (\ref{bc9})-(\ref{bc10}) has one and only one solution $v$, for $v-\uvl$ and $g_l(\uvl)$ sufficiently small.
\end{theorem}

\begin{remark}\label{rem7}
In this result we exclude the case where the boundary is characteristic {\it i.e.} the case where one of the $\lambda_k$ is equal to $0$. This case cannot be dealt with at this level of generality. On the other hand, wall boundary conditions belong to this category. They can be discussed and handled directly on the physical system under consideration. In this section we show how to do it for the NSWE equations (see Paragraph \ref{sec:wallB}). Moreover, the treatment of wall boundaries of compressible Euler equations and some two-phase systems \cite{Dias2008a, Dias2008, Dias2008b, Dutykh2007a} can be done in a similar way.
\end{remark}

\begin{remark}\label{pratique}
In practice, (\ref{bc9})-(\ref{bc10}) are written in a parametric way. We have a set of $m$ physical variables $w$ ({\it e.g.} pressure,
densities, velocities,\ldots) and we look for $w$ satisfying:
\begin{equation}\label{bc14}
        g_l(w)=0\,,\quad l=1,\ldots,p\,,
\end{equation}
\begin{equation}\label{bc15}
        \tilde l_k(\uvl)\cdot\Phi=\tilde l_k(\uvl)\cdot(F(w^n_K)\cdot \n_K)\,,
\end{equation}
\begin{equation}\label{bc16}
        \Phi=F(w)\cdot \n_K\,,
\end{equation}
and then we take:
\begin{equation}\label{bc17}
        \Phi(w^n_K,K,\partial\Om)=\Phi\,.
\end{equation}
The system (\ref{bc14})-(\ref{bc15})-(\ref{bc16}) is then solved by Newton's method.
\end{remark}

\subsubsection{Impermeable boundary}\label{sec:wallB}

Consider the case of a rigid wall boundary
\begin{equation}\label{eq:rigidwall}
  \u(\x,t)\cdot\n = 0, \quad \x\in\partial\Omega,
\end{equation}
and the hyperbolic system (\ref{eq:gov1}), (\ref{eq:gov2}). The flux $\Phi$ that we have to determine on the boundary $\partial\Omega$ has the following form if we take into account (\ref{eq:rigidwall}):
\begin{equation}\label{eq:bFlux}
  \Phi =  \left.\bigl(\F\cdot\n\bigr)\right|_{\partial\Omega} = 
  \begin{pmatrix}
    0 \\ \frac{g}{2}H^2 n_x \\ \frac{g}{2}H^2 n_y
  \end{pmatrix}.
\end{equation}
Thus, we have to determine $\frac{g}{2}H^2$ on the boundary $\partial\Omega$. For this purpose we use a complementary numerical boundary condition as explained above:
\begin{equation}\label{eq:numericalBcond}
  l_3(\w_K)\cdot\Phi = l_3(\w_K)\cdot\F_n (\w_K),
\end{equation}
where $l_3$ is the left eigenvector corresponding to the positive eigenvalue $\lambda_3 = u_n + c = c > 0$. Solving equation (\ref{eq:numericalBcond}) leads to the following value of the unknown component:
\begin{equation*}
  \left.\frac{g}{2}H^2\right|_{\partial\Omega} = 
  \left.\bigl(cHu_n + \frac{g}{2}H^2\bigr)\right|_{K},
\end{equation*}
which determines completely the boundary flux (\ref{eq:bFlux}).

\subsubsection{Generating boundary}

Now let us consider a boundary where the total water depth is prescribed:
\begin{equation*}
  \left.H\right|_{\partial\Omega} = H_0 (\x_s, t) > 0, \quad \x_s \in \partial\Omega.
\end{equation*}
Taking into account this information, the flux $\Phi$ to be determined has the following form:
\begin{equation}\label{eq:generB}
  \Phi = \left.\bigl(\F\cdot\n\bigr)\right|_{\partial\Omega} = 
  \begin{pmatrix}
    H_0 u_n \\
    H_0 u u_n + \frac{g}{2}H_0^2 n_x \\
    H_0 v u_n + \frac{g}{2}H_0^2 n_y \\
  \end{pmatrix}.
\end{equation}
Hence, we have to find $u$ and $v$ on the generating boundary $\partial\Omega$. The normal velocity will be immediately deduced from this information $u_n := un_x + vn_y$.

Throughout this section we have assumed that the flow is ``subsonic'', i.e. $\abs{\u\cdot\n} \leq c$. We could also consider the ``supersonic'' case, but physically this situation is rather exotic. Henceforth, we have one negative eigenvalue $\lambda_1 = u_n - c$, one positive $\lambda_3 = u_n + c$ and $\lambda_2 = u_n$ can be in principle of any sign. Thus we have to consider two cases: $u_n < 0$ and $u_n \geq 0$. In the first case we need a supplementary physical condition (on the tangential velocity to the boundary), in the second one we use a supplementary numerical condition:
\begin{equation*}
  l_2(\w_K)\cdot\Phi = l_2(\w_K)\cdot\F_n (\w_K).
\end{equation*}
Both lead to the same conclusion: $\left.u_\tau\right|_{\partial\Omega} = \left.u_\tau\right|_{K}$, where $u_\tau := un_y - vn_x$ is the tangential velocity. Computations similar to the previous section \ref{sec:wallB} lead to
\begin{equation*}
  \left.u_n\right|_{\partial\Omega}=\frac{Hcu_n + \frac{g}{2}(H^2 - H_0^2)}{H_0 c}, \quad
  \left.u\right|_{\partial\Omega}=\left.u_n\right|_{\partial\Omega}n_x+u_\tau n_y, \quad
  \left.v\right|_{\partial\Omega}=\left.u_n\right|_{\partial\Omega}n_y-u_\tau n_x.
\end{equation*}
Substituting these expressions into (\ref{eq:generB}) gives the boundary flux $\Phi$.


\section{Numerical results}\label{sec:results}

Two kinds of numerical tests are presented. The first kind is a comparison with analytical solutions (or approximate analytical solutions): sections \ref{sub:convergence}, \ref{sub:catalina1}, \ref{sub:catalina3}. This allows us to test the correctness and precision of the numerical scheme. The second kind is a comparison with results from laboratory experiments: section \ref{sub:catalina2}.
This allows us to test the capacity of the code to reproduce actual events, and in particular to assess the validity of the nonlinear shallow-water equations for tsunami modeling.

\subsection{Convergence test}
\label{sub:convergence} 

We begin the presentation of numerical tests by the simplest one -- convergence test. We show the accuracy of the MUSCL scheme implementation. In order to do it, we solve numerically the following scalar linear advection equation
\begin{equation*}
  \pd{w}{t} + \u_0\cdot\grad w = 0, \quad \u_0\in\R^2
\end{equation*}
with smooth\footnote{We intentionally choose a smooth initial condition since the discontinuities can decrease the overall accuracy of the scheme.} initial conditions. Moreover, it has almost compact support in order to reduce the influence of boundary
conditions. It is obvious that this equation will just translate the initial form in the direction $\u_0$. So, we have an analytical solution which can be used to quantify the numerical method error. On the other hand, to measure the convergence rate, we constructed a sequence of refined meshes.

\begin{figure}
  \centering
  \includegraphics[width=0.59\textwidth]{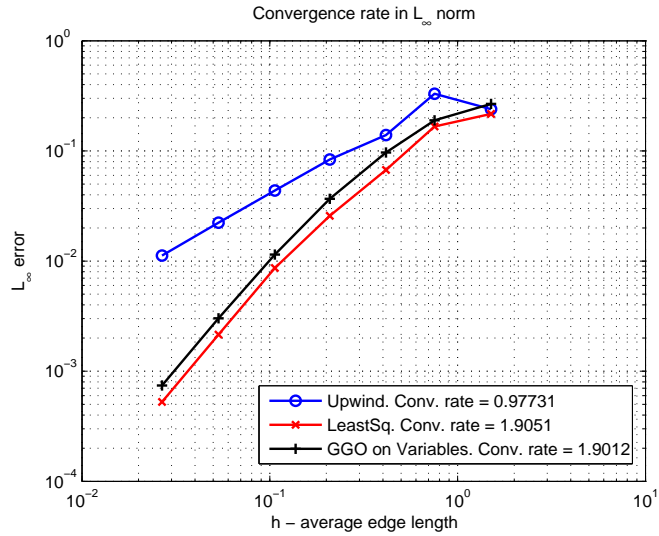}
  \caption{Error of the numerical method in $L_\infty$ norm.}
  \label{fig:convlinf}
\end{figure}

\figurename~\ref{fig:convlinf} shows the error of the numerical method in $L_\infty$ norm as a function of the mesh characteristic size. The slope of these curves represents an approximation to the theoretical convergence rate. On this plot, the blue curve corresponds to the first order upwind scheme while the other two (red and black) correspond to the MUSCL scheme with least-squares (see Section \ref{sec:lsq}) and Green-Gauss (see Section \ref{sec:ggo}) gradient reconstruction procedures respectively. One can see that the blue curve slope is equal approximatively to $0.97$ which means first order convergence. The other two curves have almost the same slope equal to $1.90$, indicating a second order convergence rate for the MUSCL scheme. In our implementation of the second-order scheme the least-squares reconstruction seems to give slightly more accurate results than the Green-Gauss procedure.

\begin{figure}
  \centering
  \includegraphics[width=0.59\textwidth]{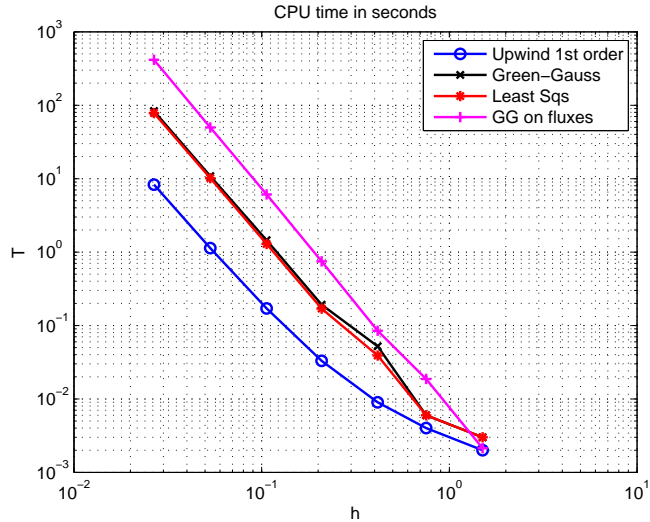}
  \caption{CPU time for different finite volume schemes.}
  \label{fig:cputime}
\end{figure}

The next figure represents the measured CPU time in seconds as a function of the mesh size. Obviously, this kind of data is extremely computer dependent but the qualitative behaviour is the same on all systems. On \figurename~\ref{fig:cputime} one can see that the ``fastest'' curve is the blue one (first order upwind scheme). Then we have two almost superimposed (black and red) curves referring to the second-order gradient reconstruction on variables. Here again one can notice that the least-squares method is slightly faster than the Green-Gauss procedure. On this figure we represented one more curve (the highest one) which corresponds to Green-Gauss gradient reconstruction on fluxes (it seems to be very natural in the context of the FVCF scheme explained in Section \ref{sec:vffc}). Our numerical tests show that this method is more expensive from the computational point of view and we decided not to choose it.

The next three sections deal with the validation of VOLNA against a set of benchmarks for tsunami modelling proposed at the Catalina 2004 workshop on long waves~; they are among the 6 benchmarks currently recommended by the United States National Oceanic and Atmospheric Administration (NOAA) for the evaluation of operational tsunami forecasting models~\cite{noaa_report}. In order to help the reproductibility and comparison of numerical results, all the following test cases make use of publicly available data\footnote{ \texttt{http://nctr.pmel.noaa.gov/benchmark/index.html}}. Although we present results for one mesh only for each benchmark, we have checked that the simulations converge as the mesh resolution increases.

\subsection{Tsunami run-up onto a plane beach}\label{cata1}
\label{sub:catalina1}

\begin{figure}[t]
\centering
%
%
%
%
    \includegraphics[width=0.6\textwidth]{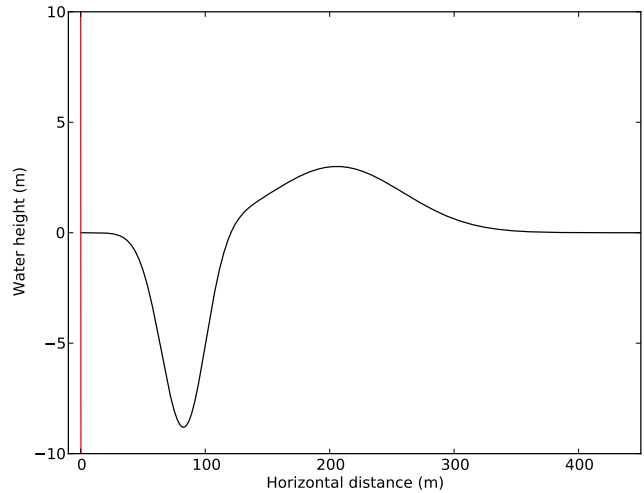}
  
%

  \caption{Catalina~1 benchmark --- bathymetry and initial free surface, at the same vertical scale.}
  \label{fig:cata1_initial}  

\end{figure}

In this test case, we look at the runup of a tsunami wave over a plane sloping beach (of slope $\tfrac{1}{10}$). The initial depression wave propagates leftwards (see figure~\ref{fig:cata1_initial}). The result of the simulation is compared to an analytical solution obtained by using the initial value problem technique of Carrier, Wu and Yeh~\cite{CWY}. Note that the computational domain is approximately 50~km long, whereas the shoreline motion scale is 1~km~; hence, we choose to refine the mesh by a factor 10 near the initial shoreline. The results presented here correspond to a resolution of 8~meters in the direction of propagation. Moreover, the initial free surface amplitude is a few meters high. Thus, due to the difference of spatial scales between the bathymetry, the initial free surface and the computational domain dimensions, the source term $gH \nabla h$ is very steep (see figure~\ref{fig:cata1_initial}), which renders the use of a well-balanced scheme mandatory. This test case is one-dimensional. Since our code is two-dimensional, we implement it using a two dimensional computational domain, with translation invariance in the transverse direction.

\readdata[xStart=-200,xEnd=800]{\numericOne}{figs/cata1/numerics/t160.txt}
\readdata[xStart=-200,xEnd=800]{\numericTwo}{figs/cata1/numerics/t175.txt}
\readdata[xStart=-200,xEnd=800]{\numericThree}{figs/cata1/numerics/t220.txt}

\begin{figure}
  \centering
  \includegraphics[width=0.7\textwidth]{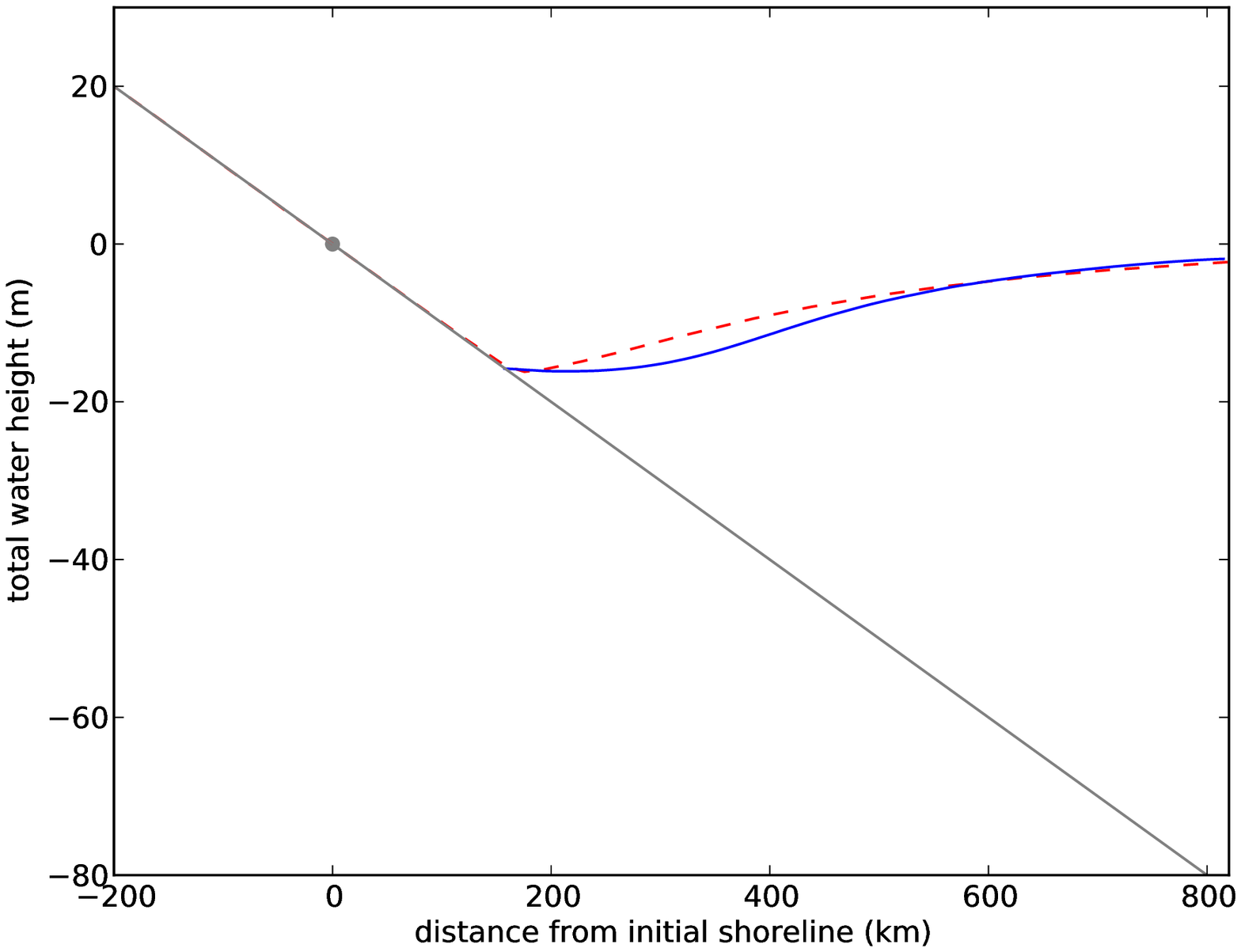}   
%
%
  
  \caption{Catalina 1 benchmark --- comparison between analytical (solid) and numerical (dashed) values for free surface at time 160 seconds. The gray line represents the beach, and the gray point the initial shoreline location.}

  \label{fig:cata1_comparison}
  
\end{figure}

\begin{figure}
  \centering
  \includegraphics[width=0.7\textwidth]{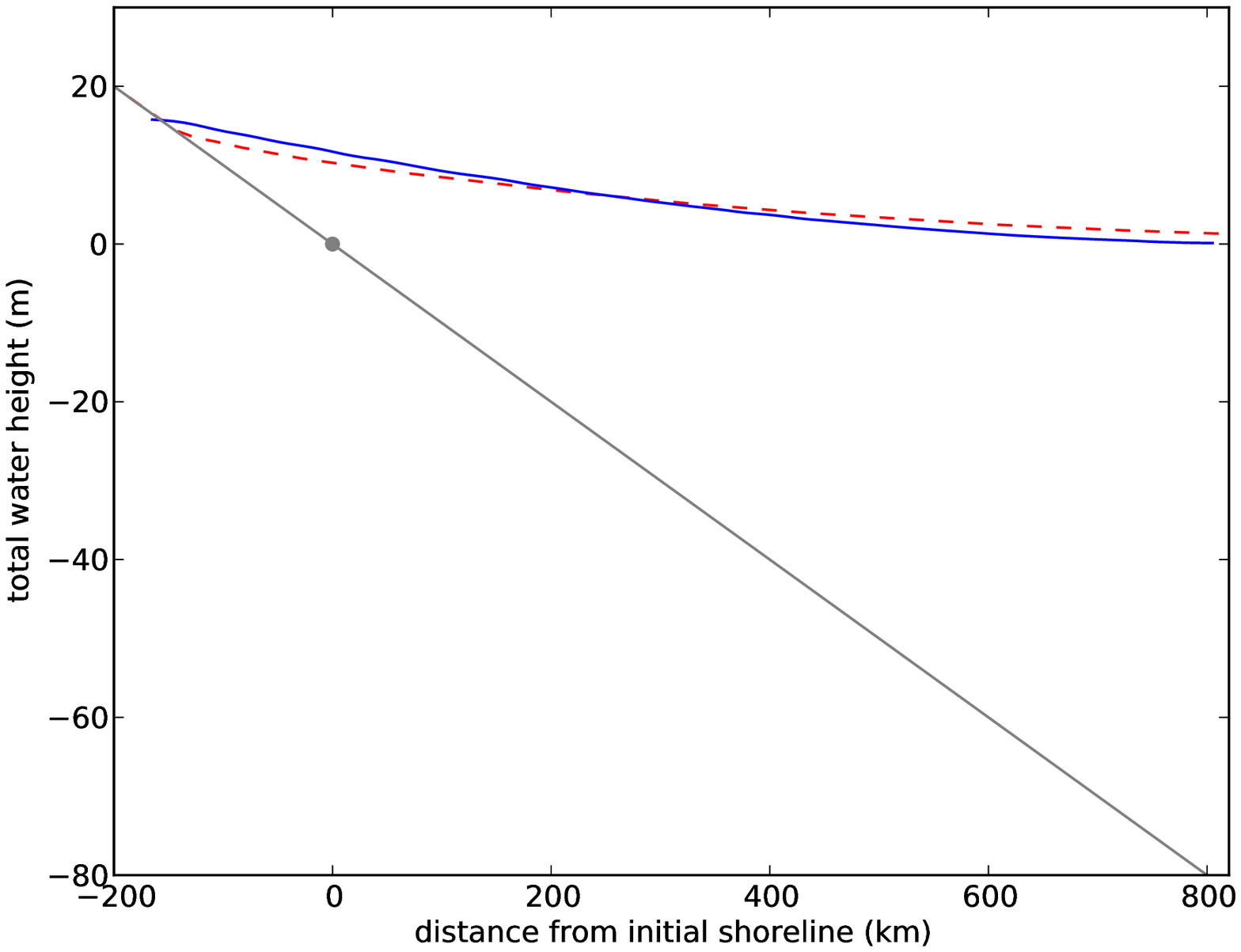}    
%
%
  
  \caption{Catalina 1 benchmark --- comparison between analytical (solid) and numerical (dashed) values for free surface at time 220 seconds. The gray line represents the beach, and the gray point the initial shoreline location.}

  \label{fig:cata1_comparison2}
  
\end{figure}

We can see on Figures \ref{fig:cata1_comparison} and \ref{fig:cata1_comparison2} that the numerical results match pretty well the approximate analytical solution, especially near the shoreline location. This ensures the accuracy of the runup algorithm presented in section~\ref{sec:runup}.

\subsection{Tsunami run-up onto a complex 3-dimensional beach}\label{cata2}
\label{sub:catalina2}

\begin{figure}[b]

  \resizebox{\linewidth}{!}{
    \psset{xAxisLabel=time ($s$), yAxisLabel=water height ($cm$),%
      xAxisLabelPos={12,-2.6}, yAxisLabelPos={-2.5,0}, tickstyle=bottom}

    \hspace{.5cm}
    
    \begin{psgraph}[Dx=5,Dy=1,Oy=-2,arrows=->,subticks=2,labelFontSize=\scriptstyle]%
      (0,-2)(25,2){8cm}{6cm}
      \fileplot[]{figs/cata2/initial_profile.txt}
    \end{psgraph}

    \hspace{1cm}
    
    \includegraphics[width=0.49\textwidth]{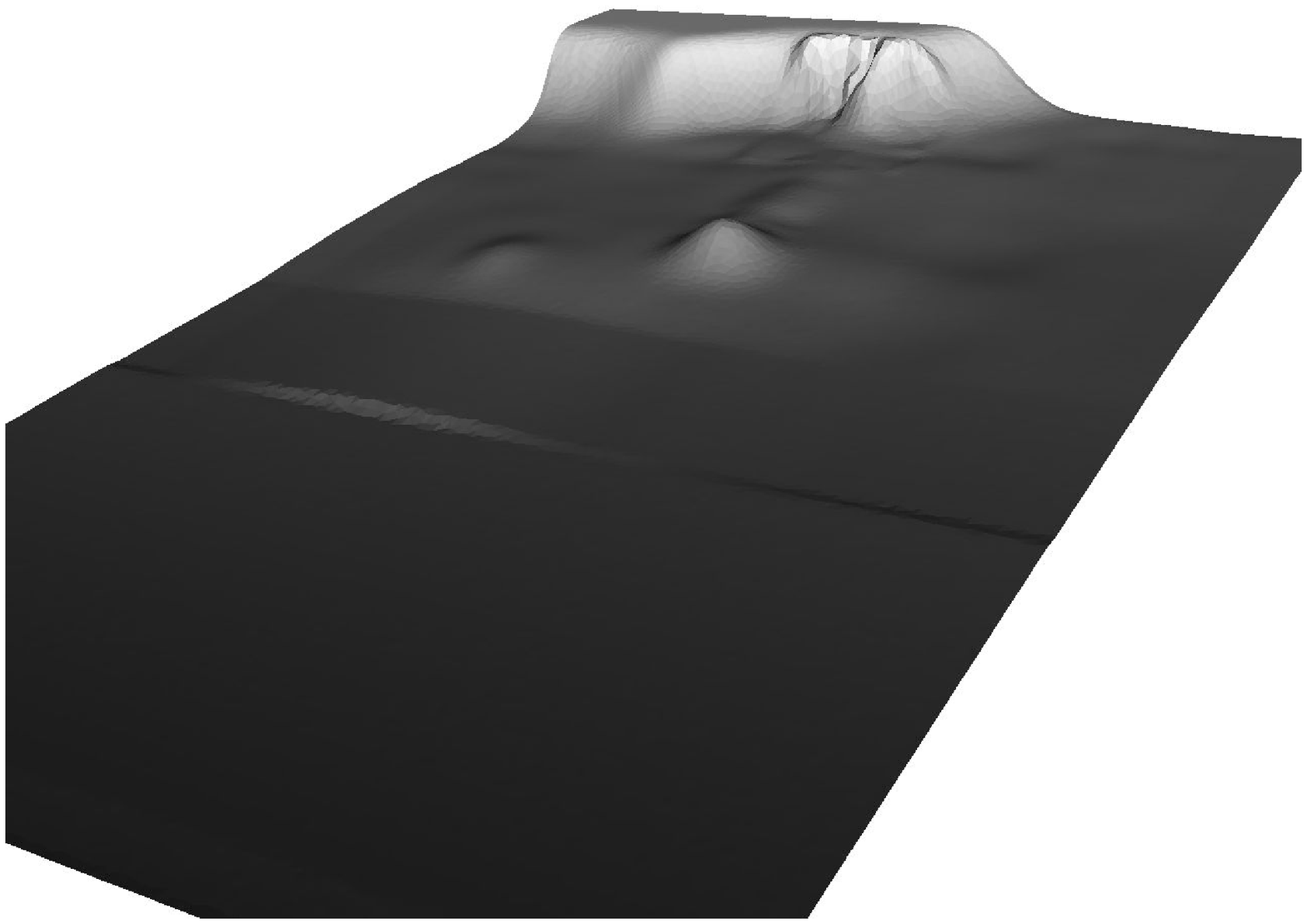}
  }

  \vspace{1cm}

  \caption{Catalina~2 benchmark~: initial free surface profile (left)~; bathymetry (right).}

  \label{fig:cata2_initial}
  
\end{figure}

\begin{figure}[t]
  \centering
        
      \psset{xAxisLabel=time ($s$), yAxisLabel=water height ($cm$),%
        xAxisLabelPos={60,-2}, yAxisLabelPos={-8,1}, tickstyle=bottom}

      \begin{psgraph}[Dx=10,Oy=-2,arrows=->,subticks=2,%
        labelFontSize=\scriptstyle](0,-2)(50,5){9cm}{5cm}
        \fileplot[]{figs/cata2/numerics/gage_5.txt}
        \fileplot[linecolor=red,linestyle=dashed]{figs/cata2/data/exp_gage_5.txt}
      \end{psgraph}
      
   \vspace{0.7cm}
   
   \caption{Catalina 2 benchmark --- comparison between numerical results and experimental data (red curve) at gage 5.}

  \label{fig:cata2_results1}
\end{figure}

\begin{figure}[t]
  \centering
        
      \psset{xAxisLabel=time ($s$), yAxisLabel=water height ($cm$),%
        xAxisLabelPos={60,-2}, yAxisLabelPos={-8,1}, tickstyle=bottom}

      \begin{psgraph}[Dx=10,Oy=-2,arrows=->,subticks=2,%
        labelFontSize=\scriptstyle](0,-2)(50,5){9cm}{5cm}
        \fileplot[]{figs/cata2/numerics/gage_7.txt}
        \fileplot[linecolor=red,linestyle=dashed]{figs/cata2/data/exp_gage_7.txt}
      \end{psgraph}
      
   \vspace{0.7cm}
   
   \caption{Catalina 2 benchmark --- comparison between numerical results and experimental data (red curve) at gage 7.}

  \label{fig:cata2_results2}
\end{figure}

\begin{figure}[t]
  \centering
        
      \psset{xAxisLabel=time ($s$), yAxisLabel=water height ($cm$),%
        xAxisLabelPos={60,-2}, yAxisLabelPos={-8,1}, tickstyle=bottom}

      \begin{psgraph}[Dx=10,Oy=-2,arrows=->,subticks=2,%
        labelFontSize=\scriptstyle](0,-2)(50,5){9cm}{5cm}
        \fileplot[]{figs/cata2/numerics/gage_9.txt}
        \fileplot[linecolor=red,linestyle=dashed]{figs/cata2/data/exp_gage_9.txt}
      \end{psgraph}
      
   \vspace{0.7cm}
   
   \caption{Catalina 2 benchmark --- comparison between numerical results and experimental data (red curve) at gage 9.}

  \label{fig:cata2_results3}
\end{figure}

This experiment reproduces at $\tfrac{1}{400}$ scale the Monai valley tsunami, which struck the Island of Okushiri (Hokkaido, Japan) in 1993, in a 205 meters long wave tank. The computational domain reproduces the last 5 meters of the wave tank. The initial incident wave offshore is given by experimental data, and fed as a time dependent boundary condition.

\begin{figure}[t]
  \centering

  \includegraphics[width=0.75\textwidth]{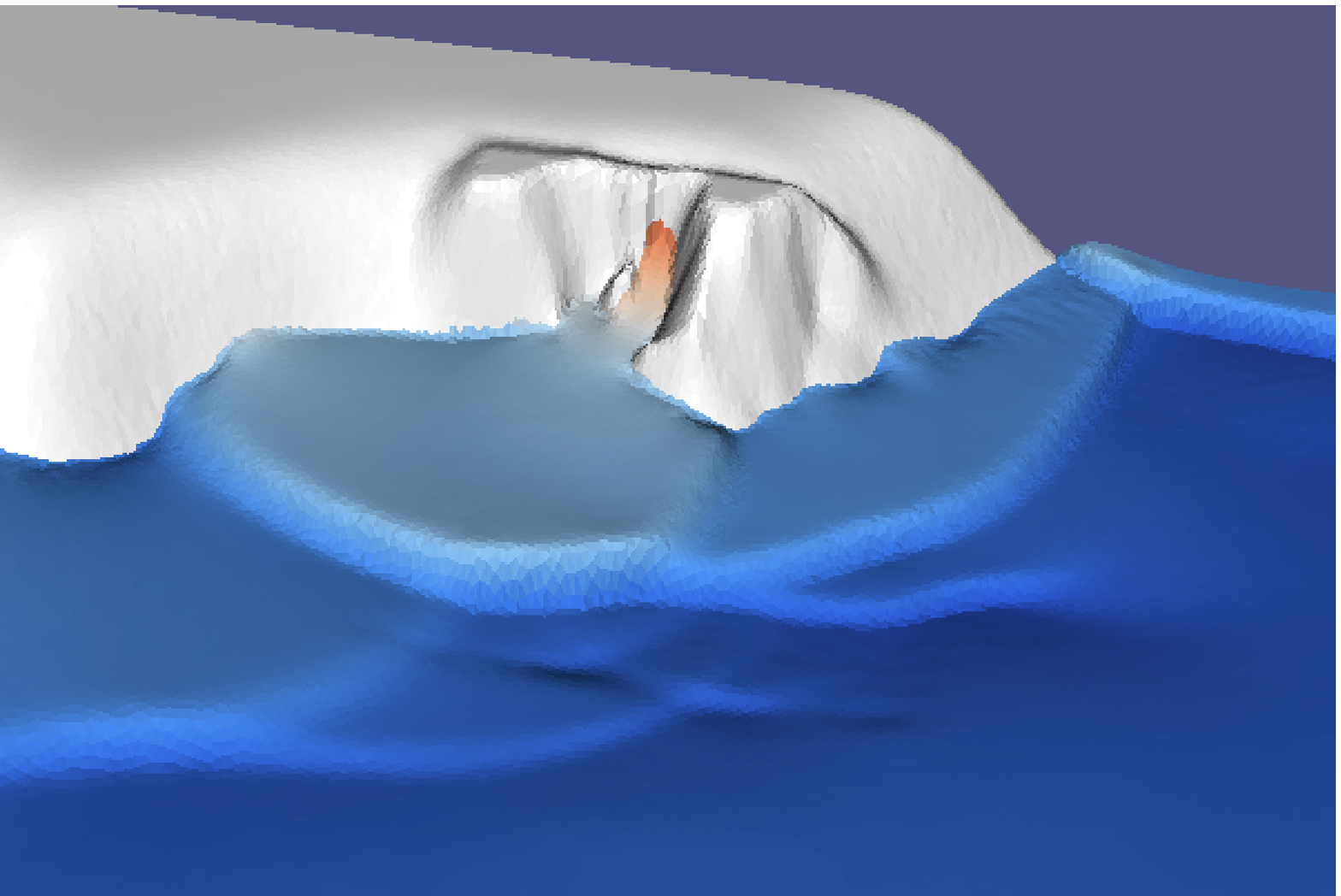}

  \caption{Maximal runup in Monai valley with a factor 3 magnification for the vertical scale.}
  \label{fig:cata2_runup}
\end{figure}

We compare the numerical results with the recorded data at three of the wave gages installed in the wave tank~: gages number 5, 7 and 9, of respective coordinates $(4.521,1.196)$, $(4.521, 1.696)$ and $(4.521, 2.196)$. This can be seen in Figures  \ref{fig:cata2_results1} -- \ref{fig:cata2_results3}. The main wave (between times 15 and 25 seconds) is very accurately described, at all gages we consider. Moreover, the maximal runup is adequately captured by the code. This value is extremely high, and occured in Monai Valley (the corresponding canyon in the experimental setup, along with the free surface at the moment of maximum elevation are shown in figure \ref{fig:cata2_runup}). We obtain a maximal runup value of 8.05 centimeters, which corresponds to 32.2 meters put back at field scale. This is remarkably close to the measured value of 31.7 meters.

Hence the numerical model is able to reproduce the laboratory scenario accurately, even without bottom friction modelling (and thus without any free parameter). In this realistic test case, the ability to refine the mesh near the zones of interest is a very nice asset.

\begin{figure}
        \centering
                \includegraphics[width=0.5\textwidth]{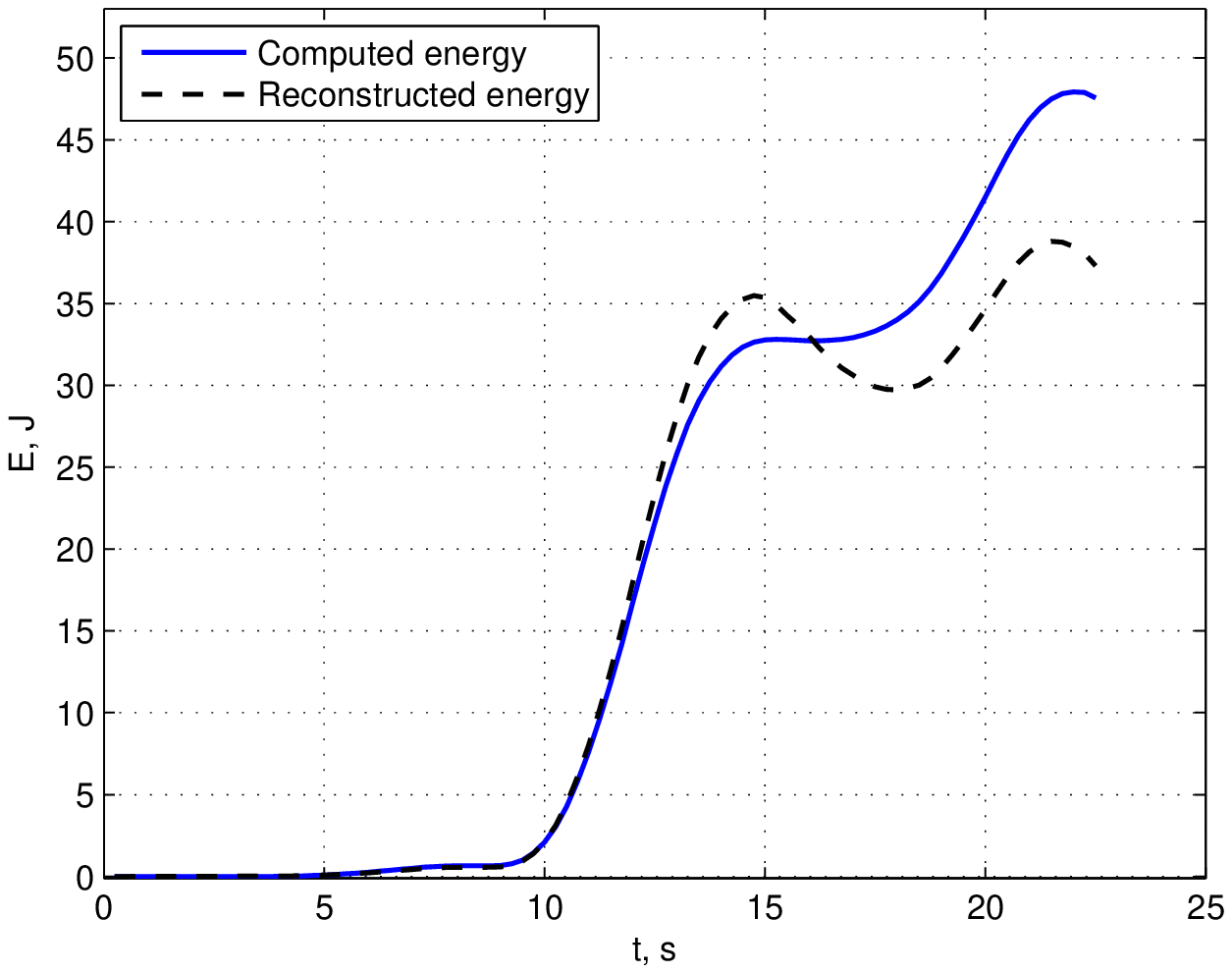}
        \caption{Computed and reconstructed wave energy profiles for Catalina 2 benchmark.}
        \label{fig:EnergyCatalina2}
\end{figure}

We performed another computation using the set-up described above. Namely, we solved equations (\ref{eq:gov1}), (\ref{eq:gov2}) completed by a conservation law for the wave energy. This system was recently proposed by Dutykh \& Dias and we refer to \cite{Dutykh2009b} for technical details and discussions. The total energy evolution is depicted on Figure~\ref{fig:EnergyCatalina2}. We represented two curves. The blue solid line corresponds to the solution of the augmented system of equations. The black dashed line refers to the total energy, estimated from conservative variables:
\begin{equation*}
  E \approx \frac12\rho H|\u|^2 + \frac12 \rho g \eta^2,
\end{equation*}
where $\rho$ is the constant fluid density and $\eta = H - h$ is the free surface elevation with respect to the undisturbed water level. In complete accordance with results reported in \cite{Dutykh2009b}, the computed wave energy is not prone to numerical diffusion and has excellent monotonicity properties. Just at the end of the simulation one can notice a little decrease in both curves. In fact, it is induced by energy losses due to the wave run-up on the beach.

\subsection{Tsunami generation and runup due to a 2-dimensional landslide}\label{cata3}
\label{sub:catalina3}


\readdata[xStart=0,xEnd=3]{\freesurfFifteen}{figs/cata3/analytical/bench3A_15.txt}

\begin{figure}[b]
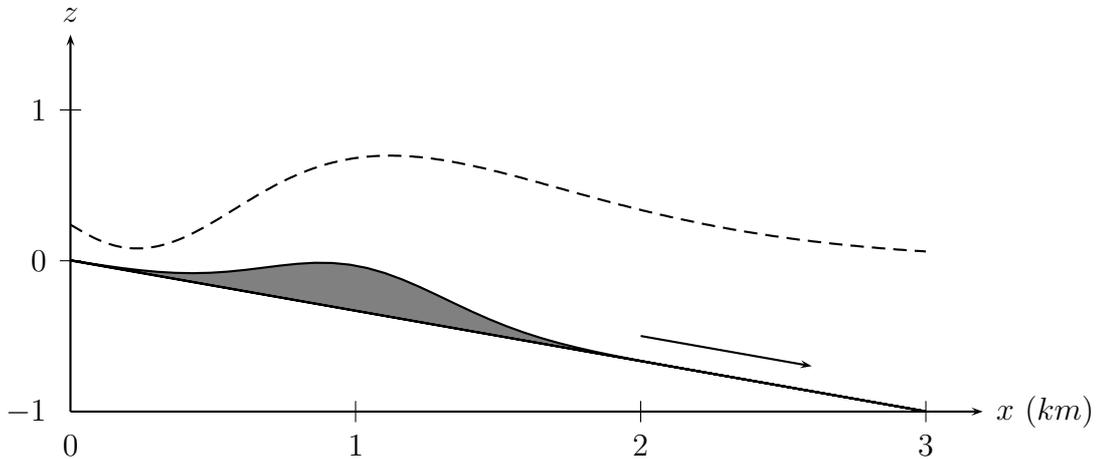

  \centering

  \psset{xAxisLabel=$x$ ($km$), yAxisLabel=$z$}
  \begin{psgraph}[Oy=-1,arrows=->](0,-1)(3.2,1.5){12cm}{5cm}
    
    \psline[arrows=->](2,-0.5)(2.6,-0.7)
    
    \pscustom[fillstyle=solid,fillcolor=gray]{
      \psplot[algebraic]{0}{3}{-x/3}
      \psplot[algebraic]{0}{3}{%
        -(x/3)+0.3*2.7^(-5*(x-1)^2)}
    }
    \listplot[linestyle=dashed]{\freesurfFifteen}
  \end{psgraph}

  \vspace{1cm}
  
  \caption{Sketch of the analytical landslide test case. The submarine mass displacement (in gray) and free surface (dashed line) have    been magnified by a factor 1000.}
  \label{fig:cata3_setup}
\end{figure}

In this test case, a translating Gaussian shaped mass, initially at the shoreline, translates rightwards and creates a wave (see figure~\ref{fig:cata3_setup}). The seafloor can be described by the following equation~:
\begin{displaymath}
  h(x,t) = H(x) - h_0(x,t),
\end{displaymath}
\begin{displaymath}
  H(x) = x\tan(\beta),\quad\mbox{and}\quad
  h_0(x,t) = \delta \exp \left [ - \biggl( 2\sqrt{\dfrac{x\mu^2}{\delta
        \tan(\beta)}}
    - \sqrt{\dfrac{g}{\delta}}\mu t \biggr)
  \right ],
\end{displaymath}
$x$ being the direction of propagation. Here, $\delta$ represent the maximum thickness of the sliding mass, $\mu$ the ratio between $\delta$ and the horizontal length of the mass, and $\beta$ the beach slope. Initially the submarine mass is partially underwater. Hence, this test case corresponds to a subaerial landslide. A sketch of this experiment can be seen in figure~\ref{fig:cata3_setup}. This benchmark is one-dimensional, but is implemented using a two-dimensional computational domain, as in section~\ref{sub:catalina1}. The numerical results we present are obtained as a one dimensional slice (which does not depend on the transverse variable).

\begin{figure}[t]
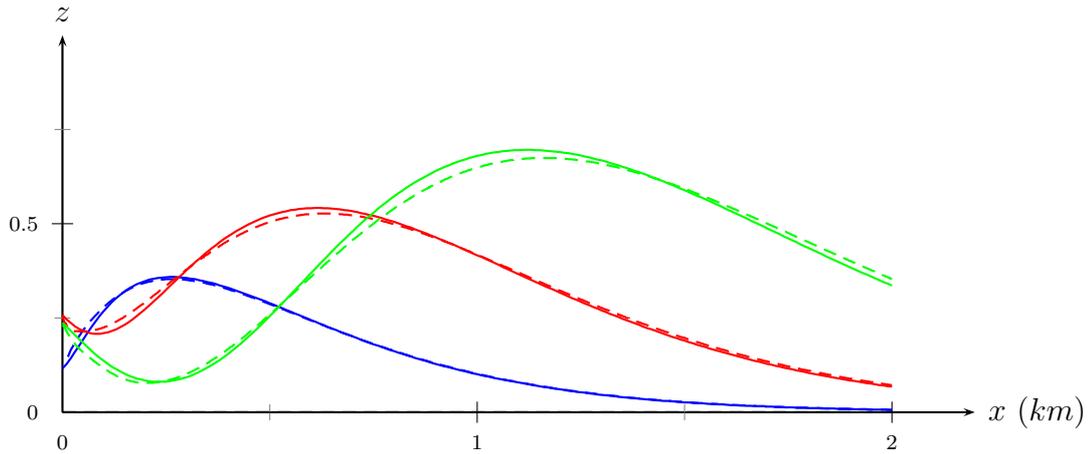

  \centering
  
  \psset{xAxisLabel=$x$ ($km$), yAxisLabel=$z$}
  \begin{psgraph}[Dy=0.5,Oy=0,arrows=->,subticks=2,
    labelFontSize=\scriptstyle](0,0)(2.2,1){12cm}{5cm}

    \psplot[linestyle=dashed]{0}{2}{0}
    \listplot[linecolor=blue]{\analyticFive}
    \listplot[linecolor=blue, linestyle=dashed]{\expFive}
    \listplot[linecolor=red]{\analyticTen}
    \listplot[linecolor=red, linestyle=dashed]{\expTen}
    \listplot[linecolor=green]{\analyticFifteen}
    \listplot[linecolor=green, linestyle=dashed]{\expFifteen}
    
  \end{psgraph}

  \vspace{1cm}
  
  \caption{Catalina 3 benchmark --- free surface profiles at three different times --- comparison between numerical results (dashed) and analytical formulas (solid lines), at times 16 (blue), 32 (red) and 48 (green) seconds.}
  \label{fig:cata3_results}
\end{figure}

The result of the numerical simulation is compared to an analytical solution, computed as an approximate solution of the linear shallow water equations with a forcing term~\cite{Liu2003}. In figure~\ref{fig:cata3_results}, we can see the comparison between the analytical and numerical wave surface profiles at times 16, 32 and 48 seconds, for $\beta = 5.7^o$, $\delta=1$ m, and $\mu=0.01$. These values ensure that we are in the linear regime of the shallow water equations (and thus that the analytical solution is a good approximation of the nonlinear equations solution). Hence, the comparison is meaningful. Good agreement is reached.

\readdata[xStart=0,xEnd=2]{\analyticFive}{figs/cata3/analytical/bench3A_05.txt}
\readdata[xStart=0,xEnd=2]{\analyticTen}{figs/cata3/analytical/bench3A_10.txt}
\readdata[xStart=0,xEnd=2]{\analyticFifteen}{figs/cata3/analytical/bench3A_15.txt}

\readdata[xStart=0,xEnd=2]{\expFive}{figs/cata3/numerics/profile_16.txt}
\readdata[xStart=0,xEnd=2]{\expTen}{figs/cata3/numerics/profile_32.txt}
\readdata[xStart=0,xEnd=2]{\expFifteen}{figs/cata3/numerics/profile_48.txt}

\begin{figure}
        \centering
         \includegraphics[width=0.49\textwidth]{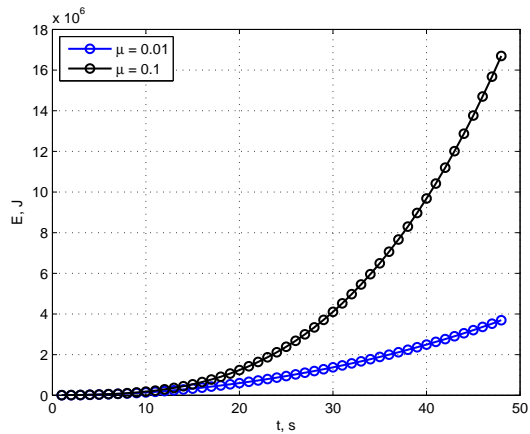}
        \caption{Energy evolution with time for the 2-dimensional landslide test-case.}
        \label{fig:catalina3a_energyAB}
\end{figure}

\begin{figure}
        \centering
         \includegraphics[width=0.49\textwidth]{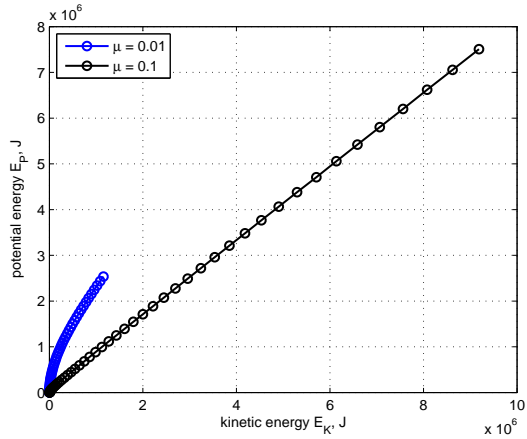}
        \caption{Trajectories in the energy phase-space $(E_k, E_p)$ for the 2D landslide test-case.}
        \label{fig:catalina3a_phaseAB}
\end{figure}

We also performed wave energy computation for this test-case. To our knowledge, the energy evolution has not been shown yet for a landslide generated wave. Computation results are presented on Figures \ref{fig:catalina3a_energyAB} and \ref{fig:catalina3a_phaseAB} for two cases $\mu = 0.01$ and $\mu = 0.1$. In the latter case the linear shallow water equations (LSWE) are not valid even on small time scales. See \cite{Liu2003} for more information.

Figure \ref{fig:catalina3a_energyAB} shows the wave energy evolution with time. On Figure \ref{fig:catalina3a_phaseAB} we represented two trajectories in the energy phase-space $(E_k, E_p)$, where $E_k$ is the kinetic energy and $E_p$ is the potential one. We would like to point out the approximate energy repartition on the black curve ($\mu = 0.1$).

\subsection{Summary}
\label{sec:summary}
Using different analytical benchmarks, we have validated all components of our code~: accuracy and order of convergence of the numerical scheme (section~\ref{sub:convergence}), run up algorithm and treatment of steep depth fields (section~\ref{sub:catalina1}), and time varying bathymetry in a conservative shallow-water framework (section~\ref{sub:catalina3}). Moreover, we have shown the capability of the code to model realistic events, using an experimental benchmark
(section~\ref{sub:catalina2}).


\section{Conclusions and perspectives}\label{sec:concl}

In the present article we provided a detailed description of the \VOLNA code, designed for complete tsunami wave modelling. Namely, we are able to simulate the whole life-cycle of a tsunami from generation to inundation. Special attention was payed to the run-up algorithm described in Section \ref{sec:runup}. The overall performance test and validation were done in Section \ref{sec:results}.

The \VOLNA code is operational and is able to run in complex and rapidly varying conditions. The use of unstructured meshes allows the handling of real coasts. Owing to the implementation of various types of boundary conditions, the code \VOLNA can be coupled to other solvers and treat exclusively the zones where the NSWE are physically relevant.

Some new results were presented concerning the energy of tsunami waves \cite{Dutykh2009b}. In particular, we show the wave energy evolution for the Catalina 2 test case (run-up on a complex 3D beach) and a landslide generated tsunami (Catalina 3 benchmark problem).

We are presently adding more physics to the \VOLNA code: dissipative effects \cite{DutykhDias2007, Dutykh2008a} (one could for example implement the dissipative terms from Bresch and Desjardins \cite{Bresch2003}) and dispersive effects. The current research activities focus on the development of robust finite volume solvers for dispersive wave models including the runup. The first attempt was made recently by D.~Dutykh \emph{et al.} (2011) \cite{Dutykh2011}. It was shown that dispersive effects might be beneficial for the description of the run-up of large, breaking waves. Finally a GPU acceleration of the \VOLNA code is underway. Preliminary results can be found in \cite{PoncetVasnier}.


\section*{Acknowledgements}

This work has been partially supported by ANR HEXECO, Project n$^o$ BLAN$07-1\_192661$, by the 2008 Framework Program for Research, Technological development and Innovation of the Cyprus Research Promotion Foundation under the Project A$\Sigma$TI / 0308(BE)/05 and by the joint french/irish Ulysses Program of the French Ministry of Foreign Affairs under the Project 23725ZA. The third author acknowledges the support from the EU project TRANSFER (Tsunami Risk ANd Strategies For the European Region) of the sixth Framework Programme under contract no. 037058. The first author acknowledges support from ANR MathOc\'ean (Project n$^o$ ANR-08-BLAN-0301-01) and from the program ``Risques gravitaires, s\'eismes'' of Cluster Environnement and the research network VOR.

The authors thank Jean-Michel Ghidaglia for very helpful discussions and his support. Without him the \VOLNA code could not have been developed in such a short time. The authors would also like to acknowledge Professor Costas Synolakis for valuable suggestions. His work on tsunami waves is an endless source of inspiration.


\bibliographystyle{alpha}
\bibliography{biblio}

\newcommand{\etalchar}[1]{$^{#1}$}
\begin{thebibliography}{CFGR{\etalchar{+}}05}

\bibitem[AB05]{Audusse2005}
E.~Audusse and M.-O. Bristeau.
\newblock A well-balanced positivity preserving "second-order" scheme for
  shallow water flows on unstructured meshes.
\newblock {\em J. Comput. Phys}, 206:311--333, 2005.

\bibitem[ABB{\etalchar{+}}04]{Audusse2004}
E.~Audusse, F.~Bouchut, O.M. Bristeau, R.~Klein, and B.~Perthame.
\newblock A fast and stable well-balanced scheme with hydrostatic
  reconstruction for shallow water flows.
\newblock {\em SIAM J. of Sc. Comp.}, 25:2050--2065, 2004.

\bibitem[AC99]{Anastasiou1999}
K.~Anastasiou and C.~T. Chan.
\newblock Solution of the 2d shallow water equations using the finite volume
  method on unstructured triangular meshes.
\newblock {\em International Journal for Numerical Methods in Fluids},
  24:1225--1245, 1999.

\bibitem[AGN05]{Alcrudo2005}
F.~Alcrudo and P.~Garcia-Navarro.
\newblock A high-resolution {G}odunov-type scheme in finite volumes for the 2d
  shallow-water equations.
\newblock {\em Int. J. Numer. Methods Fluids}, 16:489--505, 2005.

\bibitem[AMS04]{Archambeau2004}
F.~Archambeau, N.~Mehitoua, and M.~Sakiz.
\newblock Code {S}aturne: A finite volume code for the computation of turbulent
  incompressible flows - industrial applications.
\newblock {\em International Journal On Finite Volumes}, 1:1--62, 2004.

\bibitem[Aud04]{Audusse2004a}
E.~Audusse.
\newblock {\em Mod\'elisation hyperbolique et analyse num\'erique pour les
  \'ecoulements en eaux peu profondes}.
\newblock PhD thesis, Universit\'e Paris {VI}, 2004.

\bibitem[Bar04]{Barthelemy2004}
E.~Barth\'{e}l\'{e}my.
\newblock Nonlinear shallow water theories for coastal waves.
\newblock {\em Surveys in Geophysics}, 25:315--337, 2004.

\bibitem[BB73]{Boris1973}
J.P. Boris and D.L. Book.
\newblock Flux corrected transport: {S}hasta, a fluid transport algorithm that
  works.
\newblock {\em J. Comp. Phys.}, 11:38--69, 1973.

\bibitem[BBMA01]{Brocchini2001}
M.~Brocchini, R.~Bernetti, A.~Mancinelli, and G.~Albertini.
\newblock An efficient solver for nearshore flows based on the {WAF} method.
\newblock {\em Coastal Engineering}, 43:105--129, 2001.

\bibitem[BCCC97]{Batten1997}
P.~Batten, N.~Clarke, Lambert C., and D.M. Causon.
\newblock On the choice of wavespeeds for the {HLLC} {R}iemann solver.
\newblock {\em SIAM J. Sci. Comput.}, 18(6):1553--1570, 1997.

\bibitem[BCL{\etalchar{+}}11]{Bonneton2011}
P.~Bonneton, F.~Chazel, D.~Lannes, F.~Marche, and M.~Tissier.
\newblock A splitting approach for the fully nonlinear and weakly dispersive
  green--naghdi model.
\newblock {\em J. Comput. Phys.}, 230:1479--1498, 2011.

\bibitem[BD03]{Bresch2003}
D.~Bresch and B.~Desjardins.
\newblock Existence of global weak solutions for a 2d viscous shallow water
  equations and convergence to the quasi-geostrophic model.
\newblock {\em Communications in Mathematical Physics}, 238:211--223, 2003.

\bibitem[BDDV98]{Bermudez1998}
A.~Bermudez, A.~Dervieux, J.-A. Desideri, and M.~E. Vazqueza.
\newblock Upwind schemes for the two-dimensional shallow water equations with
  variable depth using unstructured meshes.
\newblock {\em Computer Methods in Applied Mechanics and Engineering},
  155:49--72, 1998.

\bibitem[BF90]{Barth1990}
T.J. Barth and P.O. Frederickson.
\newblock Higher order solution of the {E}uler equations on unstructured grids
  using quadratic reconstruction.
\newblock {\em AIAA}, 90-0013, 1990.

\bibitem[BJ89]{Barth1989}
T.J. Barth and D.C. Jespersen.
\newblock The design and application of upwind schemes on unstructured meshes.
\newblock {\em AIAA}, 0366, 1989.

\bibitem[BO04]{Barth2004}
T.J. Barth and M.~Ohlberger.
\newblock {\em Encyclopedia of Computational Mechanics, Volume 1,
  Fundamentals}, chapter Finite Volume Methods: Foundation and Analysis.
\newblock John Wiley and Sons, Ltd, 2004.

\bibitem[BQ06]{Benkhaldoun2006}
F.~Benkhaldoun and L.~Quivy.
\newblock A non homogeneous {R}iemann solver for shallow water and two phase
  flows.
\newblock {\em Flow, Turbulence and Combustion}, 76:391--402, 2006.

\bibitem[Cas90]{Casulli1990}
V.~Casulli.
\newblock Semi-implicit finite difference methods for the two-dimensional
  shallow water equation.
\newblock {\em Journal of Computational Physics}, 86:56--74, 1990.

\bibitem[CBB06]{CBB1}
R.~Cienfuegos, E.~Barthelemy, and P.~Bonneton.
\newblock A fourth-order compact finite volume scheme for fully nonlinear and
  weakly dispersive {B}oussinesq-type equations. {P}art {I}: {M}odel
  development and analysis.
\newblock {\em Int. J. Numer. Meth. Fluids}, 51:1217--1253, 2006.

\bibitem[CBB07]{CBB2}
R.~Cienfuegos, E.~Barthelemy, and P.~Bonneton.
\newblock A fourth-order compact finite volume scheme for fully nonlinear and
  weakly dispersive {B}oussinesq-type equations. {P}art {II}: {B}oundary
  conditions and model validation.
\newblock {\em Int. J. Numer. Meth. Fluids}, 53:1423--1455, 2007.

\bibitem[CC11]{Carter2011}
J.~D. Carter and R.~Cienfuegos.
\newblock The kinematics and stability of solitary and cnoidal wave solutions
  of the serre equations.
\newblock {\em Eur. J. Mech. B/Fluids}, 30:259--268, 2011.

\bibitem[CD11]{Clamond2009}
D.~Clamond and D.~Dutykh.
\newblock Practical use of variational principles for modeling water waves.
\newblock {\em Submitted}, \url{http://arxiv.org/abs/1002.3019/}, 2011.

\bibitem[CFGR{\etalchar{+}}05]{Castro2005}
M.J. Castro, A.M. Ferreiro, J.A. Garcia-Rodriguez, J.M. Gonzalez-Vida,
  J.~Macias, C.~Pares, and M.E. Vazquez-Cendon.
\newblock The numerical treatment of wet/dry fronts in shallow flows:
  Application to one-layer and two-layer systems.
\newblock {\em Mathematical and Computer Modelling}, 42:419--439, 2005.

\bibitem[CG00]{Cortes2000}
J.~Cortes and J.-M. Ghidaglia.
\newblock Upwinding at low cost for complex models and flux schemes.
\newblock In {\em Trends in Numerical and Physical Modeling for Industrial
  Multiphase Flows}, 2000.

\bibitem[CIM{\etalchar{+}}00]{Causon2000}
D.M. Causon, D.M. Ingram, C.G. Mingham, G.~Yang, and R.V. Pearson.
\newblock Calculation of shallow water flows using a cartesian cut cell
  approach.
\newblock {\em Advances in Water Resources}, 23:545--562, 2000.

\bibitem[CLM10]{ChazelLannes2010}
F.~Chazel, D.~Lannes, and F.~Marche.
\newblock Numerical simulation of strongly nonlinear and dispersive waves using
  a {G}reen--{N}aghdi model.
\newblock {\em J. Sci. Comput.}, 2010.

\bibitem[CLS04]{Cockburn2004}
B.~Cockburn, F.~Li, and C.-W. Shu.
\newblock Locally divergence-free discontinuous {G}alerkin methods for the
  {M}axwell equations.
\newblock {\em Journal of Computational Physics}, 194:588--610, 2004.

\bibitem[CP92]{Cartwright1992}
J.~H.~E. Cartwright and O.~Piro.
\newblock The dynamics of {R}unge--{K}utta methods.
\newblock {\em Int. J. Bifurcation and Chaos}, 2:427--449, 1992.

\bibitem[CWY03]{CWY}
G.~F. Carrier, T.~T. Wu, and H.~Yeh.
\newblock Tsunami run-up and draw-down on a plane beach.
\newblock {\em J. Fluid Mech.}, 475:79--99, 2003.

\bibitem[Dav88]{Davis1988}
S.F. Davis.
\newblock Simplified second-order {G}odunov-type methods.
\newblock {\em SIAM J. Sci. Statist. Comput.}, 9:445--473, 1988.

\bibitem[DCMM11]{Dutykh2011a}
D.~Dutykh, D.~Clamond, P.~Milewski, and D.~Mitsotakis.
\newblock An implicit-explicit finite volume scheme for fully nonlinear {S}erre
  equations.
\newblock {\em International Journal On Finite Volumes}, Submitted, 2011.

\bibitem[DD07a]{Dutykh2007}
D.~Dutykh and F.~Dias.
\newblock Dissipative {B}oussinesq equations.
\newblock {\em C. R. Mecanique}, 335:559--583, 2007.

\bibitem[DD07b]{DutykhDias2007}
D.~Dutykh and F.~Dias.
\newblock Viscous potential free-surface flows in a fluid layer of finite
  depth.
\newblock {\em C. R. Acad. Sci. Paris, Ser. I}, 345:113--118, 2007.

\bibitem[DD07c]{Dutykh2006}
D.~Dutykh and F.~Dias.
\newblock Water waves generated by a moving bottom.
\newblock In Anjan Kundu, editor, {\em Tsunami and Nonlinear waves}. Springer
  Verlag (Geo Sc.), 2007.

\bibitem[DD09a]{Dutykh2009b}
D.~Dutykh and F.~Dias.
\newblock Energy of tsunami waves generated by bottom motion.
\newblock {\em Proc. R. Soc. A}, 465:725--744, 2009.

\bibitem[DD09b]{Dutykh2007b}
D.~Dutykh and F.~Dias.
\newblock Tsunami generation by dynamic displacement of sea bed due to dip-slip
  faulting.
\newblock {\em Mathematics and Computers in Simulation}, 80(4):837--848, 2009.

\bibitem[DD10]{Dutykh2008}
D.~Dutykh and F.~Dias.
\newblock Influence of sedimentary layering on tsunami generation.
\newblock {\em Computer Methods in Applied Mechanics and Engineering},
  199(21-22):1268--1275, 2010.

\bibitem[DDG08a]{Dias2008b}
F.~Dias, D.~Dutykh, and J.-M. Ghiadaglia.
\newblock A compressible two-fluid model for the finite volume simulation of
  violent aerated flows. {A}nalytical properties and numerical results.
\newblock Research report, CMLA, ENS de Cachan, 2008.

\bibitem[DDG08b]{Dias2008a}
F.~Dias, D.~Dutykh, and J.-M. Ghidaglia.
\newblock Simulation of free surface compressible flows via a two fluid model.
\newblock In {\em Proceedings of OMAE2008 27th International Conference on
  Offshore Mechanics and Arctic Engineering, June 15-20, 2008, Estoril,
  Portugal}, 2008.

\bibitem[DDG10]{Dias2008}
F.~Dias, D.~Dutykh, and J.-M. Ghidaglia.
\newblock A two-fluid model for violent aerated flows.
\newblock {\em Comput. \& Fluids}, 39(2):283--293, 2010.

\bibitem[DDK06]{ddk}
D.~Dutykh, F.~Dias, and Y.~Kervella.
\newblock Linear theory of wave generation by a moving bottom.
\newblock {\em C. R. Acad. Sci. Paris, Ser. I}, 343:499--504, 2006.

\bibitem[DKK08]{DeKaKa}
A.~I. Delis, M.~Kazolea, and N.~A. Kampanis.
\newblock A robust high-resolution finite volume scheme for the simulation of
  long waves over complex domains.
\newblock {\em Int. J. Numer. Meth. Fluids}, 56:419--452, 2008.

\bibitem[DKM11]{Dutykh2011}
D.~Dutykh, T.~Katsaounis, and D.~Mitsotakis.
\newblock Dispersive wave runup on non-uniform shores.
\newblock In {\em The International Symposium on Finite Volumes for Complex
  Applications 6}. \url{http://hal.archives-ouvertes.fr/hal-00553762/}, 2011.

\bibitem[DM10]{Dias2010}
F.~Dias and P.~Milewski.
\newblock On the fully-nonlinear shallow-water generalized {S}erre equations.
\newblock {\em Physics Letters A}, 374(8):1049--1053, 2010.

\bibitem[DMCS10]{Dutykh2010d}
D.~Dutykh, D.~Mitsotakis, L.~Chubarov, and Yu. Shokin.
\newblock Horizontal displacements contribution to tsunami wave energy balance.
\newblock {\em Submitted}, \url{http://hal.archives-ouvertes.fr/hal-00530999/},
  2010.

\bibitem[DMGD10]{Dutykh2010a}
D.~Dutykh, D.~Mitsotakis, X.~Gardeil, and F~Dias.
\newblock On the use of finite fault solution for tsunami generation problem.
\newblock {\em Submitted}, \url{http://arxiv.org/abs/1008.2742}, 2010.

\bibitem[DT07]{Tkalich2007}
M.H. Dao and P.~Tkalich.
\newblock Tsunami propagation modelling - a sensitivity study.
\newblock {\em Nat. Hazards Earth Syst. Sci.}, 7:741--754, 2007.

\bibitem[Dut07]{Dutykh2007a}
D.~Dutykh.
\newblock {\em Mathematical modelling of tsunami waves}.
\newblock PhD thesis, \'{E}cole {N}ormale {S}up\'{e}rieure de {C}achan,
  December 2007.

\bibitem[Dut09]{Dutykh2008a}
D.~Dutykh.
\newblock Visco-potential free-surface flows and long wave modelling.
\newblock {\em Eur. J. Mech. B/Fluids}, 28:430--443, 2009.

\bibitem[EMRS91]{Einfeldt1991}
B.~Einfeldt, C.D. Munz, P.L. Roe, and B.~Sjogreen.
\newblock On {G}odunov-type methods near low densities.
\newblock {\em J. Comput. Phys.}, 92:273--295, 1991.

\bibitem[FT95]{Fraccarollo1995}
L.~Fraccarollo and E.F. Toro.
\newblock Experimental and numerical assessment of the shallow water model for
  two-dimensional dam break type problems.
\newblock {\em Journal of Hydraulic Research}, 33:843--864, 1995.

\bibitem[Geo06]{George2006}
D.L. George.
\newblock {\em Finite Volume Methods and Adaptive Refinement for Tsunami
  Propagation and Inundation}.
\newblock PhD thesis, Department of Applied Mathematics, {U}niversity of
  {W}ashington, {S}eattle, 2006.

\bibitem[Geo08]{George2008}
D.L. George.
\newblock Augmented {R}iemann solvers for the shallow water equations over
  variable topography with steady states and inundation.
\newblock {\em J. Comput. Phys.}, 227:3089--3113, 2008.

\bibitem[Ghi95]{Ghidaglia1995}
J.-M. Ghidaglia.
\newblock Une approche volumes finis pour la résolution des systèmes
  hyperboliques de lois de conservation, note.
\newblock Technical report, Département Transferts Thermiques et Aérodynamique,
  Direction des Etudes et Recherches, Electricité de France, HT-30/95/015/A,
  1995.

\bibitem[Ghi98]{Ghidaglia1998}
J.-M. Ghidaglia.
\newblock Flux schemes for solving nonlinear systems of conservation laws.
\newblock In J.J. Chattot and M.~Hafez, editors, {\em Proceedings of the
  meeting in honor of P.L. Roe}, Arcachon, July 1998.

\bibitem[GHS03]{Gallouet2003}
T.~Gallouet, J.-M. H\'erard, and N.~Seguin.
\newblock Some approximate {G}odunov schemes to compute shallow-water equations
  with topography.
\newblock {\em Computers \& Fluids}, 32:479--513, 2003.

\bibitem[GKC96]{Ghidaglia1996}
J.-M. Ghidaglia, A.~Kumbaro, and G.~Le Coq.
\newblock Une m\'{e}thode volumes-finis \`{a} flux caract\'{e}ristiques pour la
  r\'{e}solution num\'{e}rique des syst\`{e}mes hyperboliques de lois de
  conservation.
\newblock {\em C. R. Acad. Sci. I}, 322:981--988, 1996.

\bibitem[GKC01]{Ghidaglia2001}
J.-M. Ghidaglia, A.~Kumbaro, and G.~Le Coq.
\newblock On the numerical solution to two fluid models via cell centered
  finite volume method.
\newblock {\em Eur. J. Mech. B/Fluids}, 20:841--867, 2001.

\bibitem[GL06]{George2006a}
D.L. George and R.J. Leveque.
\newblock Finite volume methods and adaptive refinement for global tsunami
  propagation and local inundation.
\newblock {\em Sci. Tsunami Hazards}, 24(5):319, 2006.

\bibitem[Gla88]{Glaister1988}
P.~Glaister.
\newblock Approximate {R}iemann solutions of the shallow water equations.
\newblock {\em Journal of Hydraulic Research}, 26:293--300, 1988.

\bibitem[GNVC00]{Garcia-Navarro2000}
P.~Garcia-Navarro and M.~E. Vazquez-Cendon.
\newblock On numerical treatment of the source terms in the shallow water
  equations.
\newblock {\em Computers \& Fluids}, 29:951--979, 2000.

\bibitem[God59]{Godunov1959}
S.K. Godunov.
\newblock A finite difference method for the numerical computation of
  discontinuous solutions of the equations of fluid dynamics.
\newblock {\em Mat. Sb.}, 47:271--290, 1959.

\bibitem[GOSI97]{Goto1997}
C.~Goto, Y.~Ogawa, N.~Shuto, and F.~Imamura.
\newblock Numerical method of tsunami simulation with the leap-frog scheme.
\newblock Technical report, UNESCO, 1997.

\bibitem[GP05]{Ghidaglia2005}
J.-M. Ghidaglia and F.~Pascal.
\newblock The normal flux method at the boundary for multidimensional finite
  volume approximations in cfd.
\newblock {\em European Journal of Mechanics B/Fluids}, 24:1--17, 2005.

\bibitem[GST01]{Gottlieb2001}
S.~Gottlieb, C.-W. Shu, and E.~Tadmor.
\newblock Strong stability-preserving high-order time discretization methods.
\newblock {\em SIAM Review}, 43:89--112, 2001.

\bibitem[GV85]{Goodman1985}
J.D. Goodman and R.J.~Le Veque.
\newblock On the accuracy of stable schemes for 2{D} conservation laws.
\newblock {\em Math. Comp.}, 45(171):15--21, 1985.

\bibitem[Har83]{Harten1983}
A.~Harten.
\newblock High resolution schemes for hyperbolic conservation laws.
\newblock {\em J. Comp. Phys.}, 49:357--393, 1983.

\bibitem[HC89]{Holmes1989}
D.G. Holmes and S.D. Connel.
\newblock Solution of the 2d {N}avier-{S}tokes equations on unstructured
  adaptive grids.
\newblock In {\em AIAA 9th Computational Fluid Dynamics Conference}, volume
  89-1932-CP, June 1989.

\bibitem[HLvL83]{Harten1983a}
A.~Harten, P.D. Lax, and B.~van Leer.
\newblock On upstream differencing and {G}odunov-type schemes for hyperbolic
  conservation laws.
\newblock {\em SIAM Review}, 25:35--61, 1983.

\bibitem[HP79]{Hibberd1979}
S.~Hibberd and D.H. Peregrine.
\newblock Surf and run-up on a beach: a uniform bore.
\newblock {\em J. Fluid Mech.}, 95:323--345, 1979.

\bibitem[IAK{\etalchar{+}}07]{Ioualalen2007}
M.~Ioualalen, J.~Asavanant, N.~Kaewbanjak, S.T. Grilli, J.T. Kirby, and
  P.~Watts.
\newblock Modeling the 26 december 2004 {I}ndian {O}cean tsunami: Case study of
  impact in {T}hailand.
\newblock {\em Journal of Geophysical Research}, 112:C07024, 2007.

\bibitem[Ima96]{Imamura1996}
F.~Imamura.
\newblock {\em Long-wave runup models}, chapter Simulation of wave-packet
  propagation along sloping beach by {TUNAMI}-code, pages 231--241.
\newblock World Scientific, 1996.

\bibitem[KCY07]{Kim2007}
D.-H. Kim, Y.-S. Cho, and Y.-K. Yi.
\newblock Propagation and run-up of nearshore tsunamis with {HLLC} approximate
  {R}iemann solver.
\newblock {\em Ocean Engineering}, 34:1164--1173, 2007.

\bibitem[KDD07]{Kervella2007}
Y.~Kervella, D.~Dutykh, and F.~Dias.
\newblock Comparison between three-dimensional linear and nonlinear tsunami
  generation models.
\newblock {\em Theor. Comput. Fluid Dyn.}, 21:245--269, 2007.

\bibitem[KMC03]{Kim2003}
S.-E. Kim, B.~Makarov, and D.~Caraeni.
\newblock A multi-dimensional linear reconstruction scheme for arbitrary
  unstructured grids.
\newblock Technical report, Fluent Inc., 2003.

\bibitem[Kol72]{Kolgan1972}
N.E. Kolgan.
\newblock Application of the minimum-derivative principle in the construction
  of finite-difference schemes for numerical analysis of discontinuous
  solutions in gas dynamics.
\newblock {\em Uchenye Zapiski TsaGI [Sci. Notes Central Inst. Aerodyn]},
  3(6):68--77, 1972.

\bibitem[Kol75]{Kolgan1975}
N.E. Kolgan.
\newblock Finite-difference schemes for computation of three dimensional
  solutions of gas dynamics and calculation of a flow over a body under an
  angle of attack.
\newblock {\em Uchenye Zapiski TsaGI [Sci. Notes Central Inst. Aerodyn]},
  6(2):1--6, 1975.

\bibitem[Kro97]{Kroner1997}
D.~Kroner.
\newblock {\em Numerical Schemes for Conservation Laws}.
\newblock Wiley, Stuttgart, 1997.

\bibitem[Lax73]{Lax1973}
P.D. Lax.
\newblock {\em Hyperbolic Systems of Conservation Laws and the Mathematical
  Theory of Shock Waves}.
\newblock SIAM, Philadelphia, Penn., 1973.

\bibitem[LB09]{Lannes2009}
D.~Lannes and P.~Bonneton.
\newblock Derivation of asymptotic two-dimensional time-dependent equations for
  surface water wave propagation.
\newblock {\em Phys. Fluids}, 21:016601, 2009.

\bibitem[LLS03]{Liu2003}
P.L.-F. Liu, P.~Lynett, and C.E. Synolakis.
\newblock Analytical solutions for forced long waves on a sloping beach.
\newblock {\em J. Fluid Mech.}, 478:101--109, 2003.

\bibitem[LR98]{LeRoux1998}
A.-Y. Le~Roux.
\newblock Riemann solvers for some hyperbolic problems with a source term.
\newblock {\em ESAIM: Proceedings}, 6:75--90, 1998.

\bibitem[LWC98]{Liu1998}
P.L.-F. Liu, S.-B. Woo, and Y.-K. Cho.
\newblock Computer programs for tsunami propagation and inundation.
\newblock Technical report, School of Civil and Environmental Engineering,
  Cornell University, 1998.

\bibitem[Lyn06]{Lynett}
P.~J. Lynett.
\newblock Nearshore wave modeling with high-order {B}oussinesq-type equations.
\newblock {\em J. Waterway, Port, Coastal, and Ocean Eng.}, 132:346--357, 2006.

\bibitem[MBFS07]{MarcheBonneton2007}
F.~Marche, P.~Bonneton, P.~Fabrie, and N.~Seguin.
\newblock Evaluation of well-balanced bore-capturing schemes for 2d wetting and
  drying processes.
\newblock {\em Int. J. Numer. Methods Fluids}, 53(5):867--894, 2007.

\bibitem[MBS03]{Madsen03}
P.~A. Madsen, H.~B. Bingham, and H.~A. Schaffer.
\newblock Boussinesq-type formulations for fully nonlinear and extremely
  dispersive water waves: derivation and analysis.
\newblock {\em Proc. R. Soc. Lond. A}, 459:1075--1104, 2003.

\bibitem[MFS08]{MFS2008}
P.A. Madsen, D.R. Fuhrman, and H.A. Sch\"{a}ffer.
\newblock On the solitary wave paradigm for tsunamis.
\newblock {\em J. Geophysical Res.}, 113:C12012, 2008.

\bibitem[MG96]{Musaferija1996}
S.~Musaferija and D.~Gosman.
\newblock Finite-volume {CFD} procedure and adaptive error control strategy for
  grids of arbitrary topology.
\newblock {\em J. Comp. Phys.}, 138:766--787, 1996.

\bibitem[MSS97]{Madsen1997}
P.A. Madsen, H.A. Sorensen, and H.A. Schaffer.
\newblock Surf zone dynamics simulated by a {B}oussinesq-type model. {P}art
  {I}. {M}odel description and cross-shore motion of regular waves.
\newblock {\em Coastal Engineering}, 32:255--287, 1997.

\bibitem[NPPN06]{Noelle2006}
S.~Noelle, N.~Pankratz, G.~Puppo, and J.R. Natvig.
\newblock Well-balanced finite volume schemes of arbitrary order of accuracy
  for shallow water flows.
\newblock {\em J. Comput. Phys.}, 213:474--499, 2006.

\bibitem[OHK97]{Ozkan-Haller1997}
H.T. Ozkan-Haller and J.T. Kirby.
\newblock A {F}ourier-{C}hebyshev collocation method for the shallow water
  equations including shoreline runup.
\newblock {\em Applied Ocean Research}, 19:21--34, 1997.

\bibitem[Osh84]{Osher1984}
S.~Osher.
\newblock Riemann solvers, the entropy condition, and difference
  approximations.
\newblock {\em SIAM J. Numer. Anal.}, 21(2):217--235, 1984.

\bibitem[PCD{\etalchar{+}}10]{PoncetCanada2010}
R.~Poncet, C.~Campbell, F.~Dias, J.~Locat, and D.~Mosher.
\newblock A study of the tsunami effects of two landslides in the st. lawrence
  estuary.
\newblock In D.C.~Mosher et~al., editor, {\em Submarine Mass Movements and
  Their Consequences}, pages 755--764. Springer Verlag, 2010.

\bibitem[Per67]{Peregrine1967}
D.~H. Peregrine.
\newblock Long waves on a beach.
\newblock {\em J. Fluid Mech.}, 27:815--827, 1967.

\bibitem[Pet91]{Peterson1991}
T.~Peterson.
\newblock A note on the convergence of the discontinuous {G}alerkin method for
  a scalar hyperbolic equation.
\newblock {\em SIAM J. Numer. Anal.}, 28(1):133--140, 1991.

\bibitem[PV11]{PoncetVasnier}
R.~Poncet and J.C. Vasnier.
\newblock Sustainable manycore parallelization of an unstructured hydrodynamic
  code using directive-based languages open{MP} and {HMPP}.
\newblock 2011.

\bibitem[Roe81]{Roe1981}
P.~L. Roe.
\newblock Approximate {R}iemann solvers, parameter vectors and difference
  schemes.
\newblock {\em J. Comput. Phys.}, 43:357--372, 1981.

\bibitem[SB06]{Syno2006}
C.E. Synolakis and E.N. Bernard.
\newblock Tsunami science before and beyond {B}oxing {D}ay 2004.
\newblock {\em Phil. Trans. R. Soc. A}, 364:2231--2265, 2006.

\bibitem[SBG04]{Sapoval2004}
B.~Sapoval, A.~Baldassarri, and A.~Gabrielli.
\newblock Self-stabilized fractality of seacoasts through damped erosion.
\newblock {\em Phys. Rev. Lett.}, 93:098501 [4 pages], 2004.

\bibitem[SBT{\etalchar{+}}07]{noaa_report}
C.E. Synolakis, E.N. Bernard, V.V. Titov, U.~Kanoglu, and F.I. Gonzalez.
\newblock Standards, criteria, and procedures for {NOAA} evaluation of tsunami
  numerical models.
\newblock Technical report, NOAA/Pacific Marine Environmental Laboratory, 2007.

\bibitem[Ser53]{Serre1953}
F.~Serre.
\newblock Contribution à l'étude des écoulements permanents et variables dans
  les canaux.
\newblock {\em La Houille blanche}, 8:374--388 \& 830--872, 1953.

\bibitem[Shu88]{Shu1988}
C.-W. Shu.
\newblock Total-variation-diminishing time discretizations.
\newblock {\em SIAM J. Sci. Statist. Comput.}, 9:1073--1084, 1988.

\bibitem[SO88]{Shu1988a}
C.-W. Shu and S.~Osher.
\newblock Efficient implementation of essentially non-oscillatory
  shock-capturing schemes.
\newblock {\em J. Comput. Phys.}, 77:439--471, 1988.

\bibitem[SR02]{Spiteri2002}
R.~J. Spiteri and S.~J. Ruuth.
\newblock A new class of optimal high-order strong-stability-preserving time
  discretization methods.
\newblock {\em SIAM Journal on Numerical Analysis}, 40:469--491, 2002.

\bibitem[Sto57]{Stoker1957}
J.J. Stoker.
\newblock {\em Water Waves: {T}he mathematical theory with applications}.
\newblock Interscience, New York, 1957.

\bibitem[Syn05]{Synolakis2005}
C.~Synolakis.
\newblock India must cooperate on tsunami warning system.
\newblock {\em Nature}, 434:17--18, 2005.

\bibitem[Tat97]{Tatehata1997}
H.~Tatehata.
\newblock {\em Perspectives on Tsunami Hazard Reduction: Observations, Theory
  and Planning}, chapter The New Tsunami Warning System of the {J}apan
  Meteorological Agency, pages 175--188.
\newblock Springer, 1997.

\bibitem[TG97]{Titov1997}
V.V. Titov and F.I. Gonz\'{a}lez.
\newblock Implementation and testing of the method of splitting tsunami
  ({MOST}) model.
\newblock Technical Report ERL PMEL-112, Pacific Marine Environmental
  Laboratory, NOAA, 1997.

\bibitem[TGB{\etalchar{+}}05]{Titov2005}
V.V. Titov, F.I. Gonzalez, E.~N. Bernard, M.C. Eble, H.O. Mofjeld, J.C. Newman,
  and A.J. Venturato.
\newblock Real-time tsunami forecasting: Challenges and solutions.
\newblock {\em Natural Hazards}, 35:41--58, 2005.

\bibitem[Tor92]{Toro1992}
E.F. Toro.
\newblock Riemann problems and the {WAF} method for solving the two-dimensional
  shallow water equations.
\newblock {\em Philosophical Transactions: Physical Sciences and Engineering},
  338:43--68, 1992.

\bibitem[TSS94]{Toro1994}
E.F. Toro, M.~Spruce, and W.~Speares.
\newblock Restoration of the contact surface in the {HLL} {R}iemann solver.
\newblock {\em Shock Waves}, 4:25--34, 1994.

\bibitem[VC99]{Vazquez-Cendon1999}
M.E. Vazquez-Cendon.
\newblock Improved treatment of source terms in upwind schemes for the shallow
  water equations in channels with irregular geometry.
\newblock {\em Journal of Computational Physics}, 148:497--526, 1999.

\bibitem[vL79]{Leer1979}
B.~van Leer.
\newblock Towards the ultimate conservative difference scheme {V}: a second
  order sequel to {G}odunov' method.
\newblock {\em J. Comput. Phys.}, 32:101--136, 1979.

\bibitem[vL06]{Leer2006}
B.~van Leer.
\newblock Upwind and high-resolution methods for compressible flow: From donor
  cell to residual-distribution schemes.
\newblock {\em Communications in Computational Physics}, 1:192--206, 2006.

\bibitem[WMC06]{Wei2006}
Y.~Wei, X.-Z. Mao, and K.F. Cheung.
\newblock Well-balanced finite-volume model for long-wave runup.
\newblock {\em Journal of Waterwave, Port, Coastal and Ocean Engineering},
  132:114--124, 2006.

\bibitem[ZCIM02]{Zhou2002}
J.G. Zhou, D.M. Causon, D.M. Ingram, and C.G. Mingham.
\newblock Numerical solutions of the shallow water equations with discontinuous
  bed topography.
\newblock {\em Int. J. Numer. Meth. Fluids}, 38:769--788, 2002.

\end{thebibliography}

\end{document}